\documentclass[a4paper,11pt]{article}
\usepackage[utf8]{inputenc}
\usepackage{tcolorbox}
\usepackage{amsmath,amsfonts,amssymb,array,graphicx,color}
\usepackage[nosort]{cite}
\usepackage[margin=1in]{geometry}
\usepackage{hyperref}
\usepackage{graphicx}
\usepackage{blkarray} 
\usepackage{comment}
\usepackage{authblk}
\usepackage{appendix}

\hypersetup{
  colorlinks=false,
  linktoc=all
}

\graphicspath{{Figures/}}
\numberwithin{equation}{section}

\newcommand{\aver}[1]{\langle #1 \rangle}
\newcommand{\bra}[1]{\langle #1|}
\newcommand{\ket}[1]{|#1 \rangle}
\newcommand{\wh}{\widehat}

\newcommand{\nn}{\nonumber}
\newcommand{\nodag}{{\phantom{\dag}}}
\newcommand{\id}{{\boldsymbol{1}}}
\newcommand{\br}[1]{[\![ #1 ]\!]}
\newcommand{\Ab}{\bar{A}}

\newcommand{\hb}{\bar{h}}
\newcommand{\mb}{\bar{m}}
\newcommand{\rb}{\bar{r}}
\newcommand{\sbar}{\bar{s}}
\newcommand{\tb}{\bar{t}}
\newcommand{\ub}{\bar{u}}
\newcommand{\pb}{{\bar{p}}}
\newcommand{\qb}{{\bar{q}}}
\newcommand{\Lb}{\bar{L}}
\newcommand{\Tb}{\overline{T}}
\newcommand{\taub}{\bar{\tau}}
\newcommand{\Vb}{\bar{V}}
\newcommand{\vb}{\bar{v}}
\newcommand{\wb}{\bar{w}}

\newcommand{\zb}{\bar{z}}

\newcommand{\partialb}{\bar{\partial}}
\newcommand{\mub}{\bar\mu}
\newcommand{\Tr}{\mathrm{Tr}}
\renewcommand{\Im}{\mathrm{Im}}
\newcommand{\OVir}{{\rm OVir}}
\newcommand{\Vir}{{\rm Vir}}
\newcommand{\Cbb}{\mathbb{C}}
\newcommand{\Nbb}{\mathbb{N}}
\newcommand{\Tbb}{\mathbb{T}}
\newcommand{\Zbb}{\mathbb{Z}}
\newcommand{\Cc}{\mathcal{C}}
\newcommand{\M}{\mathcal{M}}
\newcommand{\Nc}{\mathcal{N}}
\newcommand{\Oc}{\mathcal{O}}
\newcommand{\Pc}{\mathcal{P}}
\newcommand{\Sc}{\mathcal{S}}
\newcommand{\Tc}{\mathcal{T}}
\newcommand{\Vc}{\mathcal{V}}

\newcommand{\repND}[3]{[#1,#2,#3]}
\newcommand{\repD}[2]{#1^{(#2)}}
\newcommand{\repT}[2]{\sigma_{#1}^{(#2)}}
\newcommand{\repTdag}[2]{\sigma_{#1}^{\dag(#2)}}

\begin{document}
\newcommand\correspondingauthor{\thanks{\textbf{arotaru@lpthe.jussieu.fr} }}

\title{The operator algebra of cyclic orbifolds}
\author{Benoit Estienne, Yacine Ikhlef, Andrei Rotaru\correspondingauthor}
\affil{Sorbonne Universit\'{e}, CNRS, Laboratoire de Physique Th\'{e}orique et Hautes \'{E}nergies, LPTHE, F-75005 Paris, France}

\maketitle

\begin{abstract}

We identify the maximal chiral algebra of conformal cyclic orbifolds. In 
terms of this extended algebra, the orbifold is a rational and diagonal 
conformal field theory, provided the mother theory itself is also 
rational and diagonal. The operator content and operator product 
expansion of the cyclic orbifolds are revisited in terms of this 
algebra. The fusion rules and fusion numbers are computed via the 
Verlinde formula. This allows one to predict which conformal blocks appear 
in a given four-point function of twisted or untwisted operators, which 
is relevant for the computation of various entanglement measures in 
one-dimensional critical systems.

\end{abstract}
\clearpage
\tableofcontents
\clearpage

\section{Introduction}
\label{sec:intro}

Two-dimensional conformal field theories (CFTs) have remained an active topic of research over the past decades, having provided important advancements and insights in many domains of theoretical physics. 
In particular, they offer a powerful framework for the calculation of entanglement entropy in one-dimensional critical systems\cite{holzhey_geometric_1994,calabrese_entanglement_2004,calabrese_entanglement_2009}. In that context, cyclic orbifolds   \cite{knizhnik1987analytic, Crnkovic89,Klemm90,borisov_systematic_1998} have played a central role in the calculation of R\'enyi entropies, which are used to study entanglement in one-dimensional (1D) critical systems  and for understanding some of the properties of theories of 3D gravity in the context of the AdS/CFT correspondence \cite{nishioka_holographic_2009}. Furthermore, cyclic orbifold CFTs are of interest for the construction of certain types of superstring theories \cite{dixon_conformal_1987,klemm_orbifolds_1990}, and constitute an active topic of research in the mathematical community as well \cite{dong_s-matrix_2021}\cite{H_hn_2022} \cite{huang_representation_2020}. 

The construction of a cyclic orbifold CFT goes as follows. One starts from a CFT model $\cal M$ (the \emph{mother CFT}),
and replicate it $N$ times, to construct the tensor product theory {\small $\mathcal{M}^{\otimes N}$}. 
The replicated theory possesses a global discrete symmetry : it is invariant under permutations of the $N$ copies. The cyclic orbifold {\small $ \mathcal{M}_N$} is obtained from {\small $\mathcal{M}^{\otimes N}$} by gauging the abelian ${\displaystyle \mathbb{Z}_N}$ subgroup of cyclic permutations.  This means enlarging the space of configurations to allow (oriented) defect lines implementing a $\mathbb{Z}_N$ domain wall. 
Such defect lines are topological, in the sense that they can be continuously deformed (keeping their extremities fixed in the case of open lines) without changing the value of the partition function and/or correlation functions. The inclusion of these defects requires the existence of new local fields called \emph{twist fields}, located at the extremities of open-ended defect lines. In turn, the state-operator correspondence implies the existence of \emph{twisted sectors}, which describe states in radial quantization around a twist field, that is in the presence of a defect line along the radial/time direction \cite{klemm_orbifolds_1990,1996_Oshikawa_Affleck,borisov_systematic_1998,Frohlich2009DEFECTLD}.

The computation of the N${}^\textrm{th}$ R\'enyi entropy for one-dimensional critical systems in the case of a single interval embedded in an infinite line (or a circle) at zero temperature is by now a standard computation \cite{holzhey_geometric_1994,calabrese_entanglement_2004,calabrese_entanglement_2009}. It boils down to the computation of the two-point correlation function of twist fields on the sphere in the ${\displaystyle \mathbb{Z}_N}$ orbifold, which is a relatively simple from the CFT point of view. In contrast, situations involving boundaries \cite{sully_bcft_2021,2022ScPP...12..141E,Estienne:2023tdw}, finite size and finite temperature \cite{Cardy:2014jwa,datta_renyi_2014,Chen2014SingleIR,2016JHEP...01..058L,mukhi_entanglement_2018,Wu2019FiniteTE,Gerbershagen2021MonodromyMF}, or simply several intervals \cite{2008JHEP...11..076C,Furukawa:2008uk,fagotti_entanglement_2010,calabrese_entanglement_2009-1,headrick_entanglement_2010,Calabrese:2010gpa,alba_entanglement_2010,Calabrese_2011,alba_entanglement_2011,Calabrese:2012ew,calabrese_entanglement_2013,coser_renyi_2014,2015JSMTE..06..021D,2016JSMTE..05.3109C,Coser:2015eba,Grava:2021yjp}  are much more complicated and to this day remain mostly unsolved for generic CFTs. The main difficulty in such cases is that the R\'enyi entropy involves either the two-point function of twist fields on non-trivial surfaces (such as the upper-half-plane or the torus), or higher-point correlation functions. In order to make progress in such cases, a more comprehensive understanding of fusion rules and their multiplicities in the ${\displaystyle \mathbb{Z}_N}$ orbifold, as well as a full characterization the conformal blocks is of crucial importance.

In this paper, we report some progress in that direction  based on the identification of the maximal chiral algebra of the cyclic orbifold. Naively, the Virasoro algebra  ${\displaystyle \OVir_N}$ is extended by $N-1$ simple currents, whose modes  ${\displaystyle L^{(r)}_{m}}$ obey the following commutation relations \cite{borisov_systematic_1998}
\begin{equation}\label{eq:OVirintro}
    \left[ L^{(r)}_m, L^{(s)}_n \right] = (m-n) L_{m+n}^{(r+s)}
    + \frac{Nc}{12} m(m^2-1) \, \delta_{m+n,0} \, \delta_{r+s,0} \,,
\end{equation}
with $r,s \in \Zbb_N$. There is however an important caveat : these $N-1$ additional currents are not local, as they have non-trivial monodromies around the twist fields.  This implies that acting with a mode ${\displaystyle L^{(r)}_{m}}$ on a local field yields a non-local one. Therefore, it is not possible to arrange \emph{local fields} into modules of the orbifold chiral algebra $\OVir_N$. The cornerstone of this paper is to identify the maximal subalgebra of (the universal enveloping algebra of) ${\displaystyle \OVir_N}$ that acts locally on fields.  It is generated by monomials of the form
\begin{equation}
\label{eq_generators_neutral_algebra}
     L_{m_1}^{(r_1)} \dots L_{m_p}^{(r_p)} \quad \text{such that} \quad r_1+\dots+r_p\equiv 0 \mod N\,,
\end{equation}
In the following we refer to this algebra as the \emph{neutral algebra} and denote it by $A_N$.

For simplicity, we restrict our investigation to mother CFTs $\mathcal{M}$ which are diagonal and rational (\emph{w.r.t.} to the action of the Virasoro algebra), thus containing finitely many primary fields $\phi_j$. One of the main results is that the cyclic orbifold $\mathcal{M}_N$ is then also rational and diagonal \emph{w.r.t.} the neutral algebra $A_N$. In particular, there are finitely many $A_N$-primary fields, i.e. annihilated by all monomials of the form \eqref{eq_generators_neutral_algebra} with $\sum_p m_p >0$. These primary operators come in three varieties: 
\begin{itemize}
\item \textit{untwisted non-diagonal} {\small $\Phi_{[j_1\dots j_N]}$}, 
\item \textit{untwisted diagonal} {\small $\Phi_{j}^{(r)}$},
\item and \textit{twisted} {\small $\sigma_j^{[k](r)}$},
\end{itemize}
 where $j$ runs over the primary operator spectrum of the mother theory $\mathcal{M}$, while the Fourier replica index $r$ and twist charge $k$ take  values in $\Zbb_N$ and $\Zbb_N^\times$, respectively. If we denote by {\small $ \chi_{[j_1\dots j_N]},    \chi_j^{(r)}$} and {\small $\chi_j^{[k](r)}$} the characters of the corresponding $A_N$-modules, the torus partition function is diagonal : 
 \begin{equation}
  Z_{\rm orb} = \sum_{J = [j_1 \dots j_N]} |\chi_{J}|^2
  + \sum_j \sum_{r=0}^{N-1} |\chi_j^{(r)}|^2
  + \sum_j \sum_{k=1}^{N-1} \sum_{r=0}^{N-1} |\chi_j^{[k](r)}|^2 \,,
\end{equation}
  As we shall argue, this classification of primary operators is particularly well suited for the determination of fusion rules, and the decomposition of correlation functions into conformal blocks.
This classification has been suggested by the work of \cite{borisov_systematic_1998}, which did not explicitly identify the neutral algebra $A_N$, but in fact determined the modular data and fusion rules for precisely the same set of operators, in the case $N=2$. In this sense, the present article generalizes the results of \cite{borisov_systematic_1998} to any prime $N$.

 Here is a summary of our results:
\begin{itemize}
\item We establish the operator content of the $\Zbb_N$ orbifold CFT built from an arbitrary rational, diagonal Virasoro CFT model. The $\Zbb_N$ invariant primary operators are identified as the primary operators of the $A_N\oplus \Ab_N$ neutral algebra. 
\item We build the modules under the chiral  algebra $A_N$, and the decomposition of the Hilbert space over these modules, for any prime $N$.
\item We use the above decomposition property to characterize orbifold conformal blocks. This allows us to decompose any four-point correlation function of local primary operators in terms of finitely many holomorphic and antiholomorphic conformal blocks, indexed by $A_N$- and $\Ab_N$-modules, respectively.
\item By studying the modular characters, and applying consistency conditions arising from unitarity and modular relations to lift degeneracies, we obtain the orbifold $\cal T$ and $\cal S$ matrices. This leads, through the Verlinde formula, to explicit expressions of the orbifold fusion numbers $\mathcal{N}_{\alpha\beta}^\gamma$. Namely :
\end{itemize}
\begin{equation}\label{eq:fusionrulesintro}
    \begin{aligned}
      & \Nc_{[i_1 \dots i_N],[j_1 \dots j_N]}^{[k_1 \dots k_N]} = \sum_{a,b=0}^{N-1} 
      N_{i_{1+a},j_{1+b}}^{k_1} \dots N_{i_{N+a},j_{N+b}}^{k_N} \,,\\
      & \Nc_{[i_1 \dots i_N],[j_1 \dots j_N]}^{k^{(r)}} = \sum_{a=0}^{N-1} N_{i_{1+a},j_1}^k \dots N_{i_{N+a},j_N}^k \,, \\
      & \Nc_{[i_1 \dots i_N],j^{(r)}}^{k^{(s)}} = N_{i_1,j}^k \dots N_{i_N,j}^k \,, \\
      & \Nc_{i^{(r)},j^{(s)}}^{k^{(t)}} = \delta_{r+s,t} \, N_{ij}^k \,. \\
      & \Nc_{i^{[p](r)} j^{[q](s)}}^{[k_1\dots k_N]} = \delta_{p+q,0} \, \sum_{\ell} \frac{S_{i\ell}S_{j\ell} \cdot S_{k_1\ell}\dots S_{k_N\ell}}{S_{1\ell}^N} \,, \\
      & \Nc_{i^{[p](r)} j^{[q](s)}}^{k^{(t)}} =  \frac{\delta_{p+q,0}}{N} 
      \sum_{\ell} \left[ \frac{S_{i\ell}S_{j\ell}S_{k\ell}^N}{S_{1\ell}^N}
        + \sum_{n=1}^{N-1} \omega^{np(r+s-t)} \frac{(P_{-n})_{i\ell}(P_{n})_{j\ell} S_{k\ell}}{S_{1\ell}} \right] \,, \\
      & \Nc_{i^{[p](r)} j^{[q](s)}}^{k^{[m](t)}} =  \frac{\delta_{p+q,m}}{N}
      \sum_{\ell} \left[
        \frac{S_{i\ell}S_{j\ell}S_{k\ell}}{S_{1\ell}^N} + \sum_{n=1}^{N-1} \omega^{n(r+s-t)}
        \frac{(P_{pn^{-1}}^\dag)_{i\ell} (P_{qn^{-1}}^\dag)_{j\ell} (P_{mn^{-1}}^\nodag)_{k\ell}}{S_{1\ell}}
        \right] \,.
  \end{aligned}
\end{equation}
Here $N_{ij}^k$ and $S_{ij}$ are, respectively, the fusion numbers and modular $S$-matrix of the mother theory, and $\omega$ stands for the first {\small N${}^\textrm{th}$} root of unity $\omega=\exp\left( 2\pi i/N\right)$. The matrices $P_n$'s, which are labelled by $n \in \Zbb_N^\times$, are also modular matrices acting on characters of the mother CFT -- see (\ref{eq:Pn}--\ref{eq:qn}), and $n^{-1}$ stands for the inverse of $n$ in $\Zbb_N^\times$. The sums over $\ell$ run over the primary operators of the mother CFT.

The paper is organized as follows. In Section~\ref{sec:background}, we provide some background on how the cyclic orbifold construction arises in the study of entanglement in 1D critical systems. Then we define, for a generic integer $N$, the basic properties of the cyclic orbifold such as twist operators, the conserved currents $T^{(r)}$, the orbifold Virasoro algebra $\OVir_N$, and the neutral algebras $A_N$. In Section~\ref{sec:op-algebra}, we restrict the discussion to $N$ prime, and describe the operator content, OPEs and conformal blocks. In Section \ref{sec:modular}, we analyse the modular properties of the torus partition function and derive the fusion numbers.  In Section~\ref{sec:examples}, we give some applications of our results for fusion rules and conformal blocks of the $\Zbb_3$ orbifold of minimal CFTs.
In Section~\ref{sec:conclusion}, we conclude with a recapitulation of our results, and comment on possible refinements and extensions. We have relegated to the Appendix the more technical proofs, to avoid congesting the logical flow of the article.

\section{Cyclic orbifolds}
\label{sec:background}

In this section, we provide some basic background on cyclic orbifolds. We shall first bring the reader to some intuition on how the cyclic orbifold arises in the study of entanglement in 1D systems, with a presentation of the replica trick, and show how orbifold correlators are related to physically relevant quantities such as R\'enyi entropies. By now, the computation of entanglement entropies in the context of conformal field theory is pretty standard, and for a more detailed exposition the reader can consult the reviews \cite{calabrese_entanglement_2009,nishioka_holographic_2009,headrick_entanglement_2010}. Following this, we define the basic features of the cyclic orbifold $\mathcal{M}_N$, such as the twist operators, the conserved currents, the symmetry algebra $\OVir_N$, and  the induction procedure.

\subsection{Motivation: Rényi entropies of quantum critical 1D systems}
\label{sec:motivation}

\begin{figure}[h!]
  \begin{center}
    \begin{tabular}{cc}
      \includegraphics{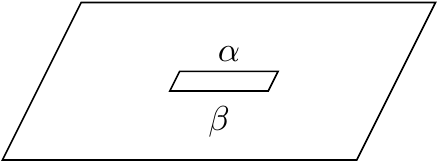} & \includegraphics{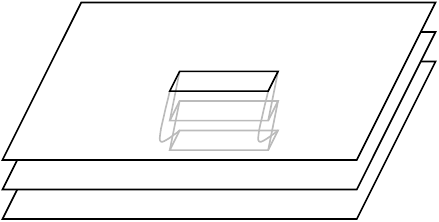} \\
      (a) & (b)
    \end{tabular}
    \caption{(a) The partition function associated to a matrix element $(\rho_A)_{\alpha \beta}$ of the ground-state density matrix $\rho_A$. (b) The partition function associated to $\Tr_A(\rho_A^N)$ for $N=3$.}
    \label{fig:cut}
  \end{center}
\end{figure}

Let us first briefly recall how R\'enyi entropies are related to partition functions on replicated surfaces \cite{holzhey_geometric_1994,calabrese_entanglement_2004}, and in turn to correlation functions of twist operators of cyclic orbifold CFTs. While it is possible to deal with systems at finite temperature and/or finite volume, for simplicity we consider a quantum critical 1D system on the infinite line $\mathbb{R}$ at zero temperature. 
We introduce a bipartition of the system into two spatial subregions, namely the interval $A=[0,\ell]$ and its complement $B = \mathbb{R} \backslash A$. The $N^{\textrm{th}}$ R\'enyi entropy (relative to this bipartition) is defined through
\begin{equation}
  S_N = \frac{1}{1-N} \log \Tr_A(\rho_A^N) \,,
  \qquad \rho_A = \frac{\Tr_B(\ket{\Psi_0}\bra{\Psi_0})}{\aver{\Psi_0|\Psi_0}} \,,
\end{equation}
where $\ket{\Psi_0}$ denotes the ground-state of the whole system, and $\rho_A$ is the reduced density matrix describing the state of the region $A$.

In the transfer-matrix (or Euclidean path integral) formalism, the quantum Hamiltonian for the physical 1D system is viewed as the anisotropic limit of the transfer matrix of a classical 2D statistical model, which we call $\M$. For any pair of states $(\alpha,\beta)$ of the subsystem $A$, the matrix element $(\rho_A)_{\alpha \beta}$ is given by the ratio 
\begin{equation}
  (\rho_A)_{\alpha\beta} = \frac{Z_{\alpha\beta}}{Z_1} \,,
\end{equation}
where $Z_1$ stands for the partition function of $\M$ on the plane, and $Z_{\alpha \beta}$ is the partition function on the plane with a cut along a segment of length $\ell$, and boundary conditions defined by $\alpha$ and $\beta$ on both sides of the cut: see Fig.~\ref{fig:cut}a. By construction, $\rho_A$ has unit trace. If $N$ is a positive integer, the quantity $\Tr_A(\rho_A^N)$ is obtained by stacking $N$ copies of the plane, and connecting them cyclically along the cut, to form the closed replicated surface $\mathcal{R}_N$ depicted in Fig.~\ref{fig:cut}.b. Thus, we have the relation
\begin{equation}
  \Tr_A(\rho_A^N) = \frac{Z_N}{Z_1^N} \,,
\end{equation}
where $Z_N$ is the partition function of the system on the replicated surface $\mathcal{R}_N$. This surface is a ramified covering of the plane, with branch points located at the extremities of the interval $A$. Geometrically, these branch points are conical singularities with an excess angle $2\pi (N-1)$. Conical singularities are \emph{not} singularities as far as the smooth (and even complex) structure is concerned\footnote{in the sense that there exists a smooth (and in fact holomorphic) coordinate chart about any branch point.}. But they are singularities in the sense of Riemannian geometry : while the metric is flat everywhere else on $\mathcal{R}_N$, there is an infinite negative curvature (proportional to the excess angle) localized at each
branch points, in accordance with the Gauss-Bonnet theorem.

\bigskip

Instead of working with replicated surfaces, we can adopt a different perspective, and replicate the degrees of freedom. To establish this idea, one can discretize the theory $\M$ and consider instead a 2D statistical model such as a classical spin model on the square lattice, with nearest-neighbour interactions. The partition function on the plane is
\begin{equation}
  Z(\M) = \sum_{\{s(i)\}} \prod_{\aver{i,j}} W[s(i),s(j)] \,,
\end{equation}
where $s(i)$ is the spin on site $i$, $\aver{i,j}$ denotes a pair of neighbouring sites, and $W$ is the Boltzmann weight which defines the local interaction.

For any positive integer $N$, we define the (lattice version of the) cyclic orbifold $\M_N$ as follows. The model $\M_N$ lives on the original surface, say the plane in the above example. At each lattice site $i$, we have $N$ independent spin variables $(s_1(i), \dots, s_N(i))$, and the nearest-neighbour interaction is given by $\prod_{a=0}^{N-1} W[s_a(i),s_a(j)]$ -- in other words, the copies $a=0, \dots, N-1$ are decoupled. The partition function thus reads
\begin{equation}
  Z(\M_N) = \sum_{\{s_a(i)\}} \prod_{\aver{i,j}} \prod_{a=1}^N W[s_a(i),s_a(j)] = Z(\mathcal{M})^N \,.
\end{equation}
Local operators are $N$-tuples of $\M$-operators. Additionally, since the model has an internal $\Zbb_N$ symmetry generated by cyclic permutations of the copies, we can have topological defects, which are oriented lines across which the local interaction is modified as $W[s_a(i),s_a(j)] \mapsto W[s_a(i),s_{a+1}(j)]$. We thus introduce the \emph{twist operators} $\sigma$ and $\sigma^\dag$, which insert the endpoints of a topological defect of this type. The construction of such defect operators is in fact pretty standard in statistical models enjoying a discrete or continuous symmetry -- this includes for instance $\Zbb_2$ disorder operators in the Ising model \cite{Kadanoff1971DeterminationOA}, and ``electric'' defects in the six-vertex model \cite{Zuber:1988he}.

Via the above construction, correlation functions of twist operators on a surface $\Sigma$ exactly correspond to partition functions of $\M$ on ramified coverings of $\Sigma$ with branch points at the position of the twists. In particular, in the above example, we have
\begin{equation}
  \aver{\sigma(0)\sigma^\dag(\ell)} = \frac{Z_N}{Z_1^N} \,,
\end{equation}
where the left-hand side is a correlator of the orbifold model $\M_N$ on the plane, and the numerator $Z_N$ on the right-hand side is, as above, the partition function of the original model $\M$ on the ramified covering $\mathcal{R}_N$ of the plane. Thus, in this simple example, we see that the R\'enyi entropy for positive integer $N$ is related to a correlation function of twist operators in the cyclic orbifold $\M$:
\begin{equation}
  S_N = \frac{1}{1-N} \log \aver{\sigma(0)\sigma^\dag(\ell)} \,.
\end{equation}
Since the original 1D quantum model is assumed to be critical and to have local interactions, the scaling limit of the lattice model $\M$ is a 2D CFT. The lattice twist operators then scale to primary operators in the cyclic orbifold CFT corresponding to $\mathcal{M}_N$, and we still denote them by $\sigma, \sigma^\dag$. Their conformal dimension is given \cite{borisov_systematic_1998} by $h_\sigma=(N-N^{-1})c/24$, where $c$ is the central charge of the mother theory: see \eqref{eq:twist_dimension}. The two-point function thus reads:
\begin{equation}
  \aver{\sigma(0)\sigma^\dag(\ell)} = \ell^{-4h_\sigma} \,,
\end{equation}
which yields the well-known result (see the seminal work \cite{calabrese_entanglement_2004})
\begin{equation}
  \label{eq:seminal_EE_scaling}
  S_N(\ell) \sim \frac{c}{6}\frac{(N+1)}{N} \log \ell \,.
\end{equation}
Several other physical situations can be treated in the framework of cyclic orbifolds.
For instance, if the subsystem $A$ is the union of $m$ disjoint intervals, then the corresponding R\'enyi entropy is related to a $2m$-point correlator of twist operators \cite{calabrese_entanglement_2009}. We  should also mention the R\'enyi entropy of a single interval in an excited primary state (instead of the ground state), which is related to the four-point correlation function
\begin{equation}
  \aver{\Phi(\infty)\sigma(1)\sigma^\dag(z,\zb)\Phi(0)} \,,
\end{equation}
where $\Phi = \phi \otimes \dots \otimes \phi$ is the corresponding replicated local primary operator in the cyclic orbifold CFT \cite{2011PhRvL.106t1601A}. Another example is the symmetry-resolved R\'enyi entropy \cite{goldstein_symmetry-resolved_2018,xavier_equipartition_2018,Calabrese:2021wvi,estienne_finite-size_2021,bonsignori_boundary_2021,Capizzi_2020}, which involves  composite twist operators $\sigma_\mu, \sigma_\mu^\dag$ obtained by fusing the bare twist operators $\sigma, \sigma^\dag$ with a family of primary operators $\phi_\mu$ associated to an internal symmetry of the original model $\M$.

\subsection{The cyclic orbifold CFT}
\label{sec:cyclic-orbifold}

With the above motivation in mind, we define cyclic orbifold CFTs as follows. Let $\M$ be any CFT  with central charge $c$ and primary operator content $\{\phi_1, \phi_2, \phi_3, \dots \}$, which we call the mother CFT. By convention, we always take $\phi_1$ to be the identity operator: $\phi_1=\id$. The cyclic orbifold CFT $\M$ for a positive integer $N$,  which is usually denoted as  
\begin{equation}
  \M_N = \M^{\otimes N} / \Zbb_N \,,
\end{equation}
is most easily defined in the path-integral formalism. If the configurations of the mother theory are described by a field $\varphi(r)$ with action $\mathcal{A}[\varphi]$, then the configurations of $\M^{\otimes N}$ are (locally) described by $N$ independent copies $(\varphi_1(r), \dots \varphi_N(r))$, with action $\mathcal{A}[\varphi_1]+\dots+\mathcal{A}[\varphi_N]$.  Furthermore, the modding out by $\mathbb{Z}_N$ means that configurations with topological defects (as described in the previous section) are to be included. In particular, on the cylinder (and on the torus,  see section \ref{sec:torus_partition_function}) this implies that one must consider all possible $\mathbb{Z}_N$ twisted boundary conditions. Accordingly, the Hilbert space (for a closed system) splits into $N$ distinct sectors labelled by the \emph{twist charge} $[k] \in \mathbb{Z}_N$, corresponding to boundary conditions twisted by the permutation of copies $a \mapsto a+k$. Via the state-operator correspondence, local fields also carry a twist charge.

In the untwisted sector $[k=0]$, the Hilbert space is simply the $N^{\textrm{th}}$ tensor product of the mother theory Hilbert space. The associated operators are called \emph{untwisted} operators and are spanned
by products of local operators acting on each copy. These are generated, under the Operator Product Expansion (OPE), by operators acting on a single copy $a$
\begin{equation}
  \phi_a = \id \otimes \dots \otimes \id \otimes \underset{(a)}{\phi} \otimes \id  \otimes \dots \otimes \id \,.
\end{equation}
By contrast, the Hilbert space in the twisted sector $[k=1]$ is in one-to-one correspondence with the mother theory Hilbert space \cite{borisov_systematic_1998}. Indeed, one can untangle the $N$ copies coupled via the twist $a \to a+1$ into a single copy, at the cost of making the system $N$ times larger. More generally, the Hilbert space in the twisted sector $[k]$ maps to the tensor product of $C_k$ copies of the mother theory Hilbert space, where $C_k$ is the number of cycles of the permutation $a \to a+k$ (in particular $C_k =1$ exactly when $k$ and $N$ are coprime).

A generic operator $\Oc^{[k]}$ with twist charge $[k]$ inserts a defect line as explained above,
and hence its monodromy relative to a diagonal operator $\phi_a$ reads:
\begin{equation}
  \phi_{a}(e^{2i\pi}z, e^{-2i\pi}\zb) \cdot \Oc^{[k]}(0)
  = \phi_{a-k}(z,\zb) \cdot \Oc^{[k]}(0) \,.
\end{equation}
being understood that upon going around the twisted field $\Oc^{[k]}(0)$, the field $\phi_a(z,\zb)$ does not encircle any other (twisted) field. 

Note that the twist charge $[k]$ is conserved under fusion, in the sense that
\begin{align}
  \aver{\Oc_1^{[k_1]}(z_1,\zb_1) \dots \Oc_n^{[k_n]}(z_n,\zb_n)}=0
  \quad \textrm{if} \quad k_1+\dots+ k_n \neq 0 \mod N \,.
\end{align}
As previously mentioned, visualizing the twisted fields as points where defect lines are inserted, with each line carrying an associated charge $[k]$, proves to be a useful approach. Under this interpretation, the charge neutrality condition in the correlator above can be understood as a topological constraint, ensuring that all defect lines must have two ends.

Of particular interest is the bare twist operator $\sigma^{[k]}$, which is simply the most relevant operator with twist charge $[k]$. Under the state-operator correspondence, it maps to the lowest energy state in the twisted sector $[k]$. In the untwisted sector this is simply the identity $\sigma^{[0]}=\id$. Moreover, to lighten the notation, we write $\sigma$ and $\sigma^{\dag}$ instead of $\sigma^{[1]}$ and  $\sigma^{[-1]}$, respectively.
The conformal dimension of the twist operators $\sigma$ and $\sigma^{\dag}$ is
\begin{equation}
  \label{eq:twist_dimension}
  h_{\sigma} = \frac{c}{24} \left(N-\frac{1}{N} \right) \,.
\end{equation}
This fundamental result, which is at the heart of the universal behaviour \eqref{eq:seminal_EE_scaling} of entanglement entropy for 1D critical systems, is surprisingly easy to establish. Indeed, it is a simple consequence of the finite size-scaling of eigenenergies in CFT.  For a periodic system of size $L$, the orbifold Hamiltonian is $\frac{2\pi}{L} (L_0 + \overline{L}_0 - Nc/12)$, as the orbifold central charge is $N c$, where $c$ is the central charge of the mother theory. Thus, the lowest energy in the sector $[k=1]$ is 
\begin{equation}
  E =  \frac{4\pi}{L} \left( h_{\sigma}  -   \frac{ N c}{24} \right)\,.
\end{equation}
since $\sigma$ is a scalar field ($h_{\sigma} = \bar{h}_{\sigma}$). But as stated above, the twisted sector $[1]$ for a system of size $L$ is nothing but the Hilbert space of the mother theory for a system of size $N L$. From this perspective the Hamiltonian is $\frac{2\pi}{NL} (L_0 + \overline{L}_0 - c/12)$, and the state with the lowest energy is the vacuum, therefore
\begin{equation}
E =   - \frac{4\pi}{NL}  \frac{ c}{24}\,.
\end{equation}
Equating these two expressions yields \eqref{eq:twist_dimension}. The argument is easy to generalize to arbitrary $k$, and one finds
\begin{equation}\label{eq:genericN_twistdimension}
  h_{\sigma^{[k]}} = \frac{c}{24} \sum_{\text{cycle } j} \left(n_{kj}-\frac{1}{n_{kj}} \right) \,.
\end{equation}
where the sum is over all cycles of the permutation $a \mapsto a+k$, and $n_{kj}$ is the length of the $j^{\textrm{th}}$ cycle. In particular, when $N$ is prime, one gets $ h_{\sigma^{[k]}}=h_{\sigma}$ for all $k \in \Zbb_N^\times$.

\subsection{Orbifold Virasoro algebra}
\label{sec:OVir}

Now that we have discussed the  splitting of the Hilbert space into twisted sectors, we can describe the action of the extended orbifold algebra.   
In the orbifold CFT $\M_N$, each copy $a$ of the mother CFT carries the components $T_a(z),\Tb_a(\zb)$ of the stress-energy tensor, with OPEs
\begin{equation}
  \begin{aligned}
    T_a(z)T_b(w) &= \delta_{ab} \left[
      \frac{c/2}{(z-w)^4} + \frac{2T_b(w)}{(z-w)^2} + \frac{\partial T_b(w)}{z-w}
      \right] + \mathrm{reg}_{z \to w} \,, \\
    \Tb_a(\zb)\Tb_b(\wb) &= \delta_{ab} \left[
      \frac{c/2}{(\zb-\wb)^4} + \frac{2\Tb_b(\wb)}{(\zb-\wb)^2} + \frac{\partialb \Tb_b(\wb)}{\zb-\wb}
      \right] + \mathrm{reg}_{\bar{z} \to \bar{w}} \,, 
  \end{aligned}
\end{equation}
where $c$ is the central charge of the mother CFT, and $\mathrm{reg}_{z \to w}$ denotes a function which is regular as $z$ tends to $w$. It turns out to be convenient to work with the discrete Fourier modes of these currents, namely
\begin{equation}
  T^{(r)}(z) = \sum_{a=0}^{N-1} \omega^{ar} \, T_a(z) \,,
  \qquad \Tb^{(r)}(\zb) = \sum_{a=0}^{N-1} \omega^{ar} \, \Tb_a(\zb) \,,
\end{equation}
where $\omega=\exp(2i\pi/N)$ and $r \in \Zbb_N$. We get the OPEs
\begin{equation} \label{eq:T(r).T(s)}
  \begin{aligned}
    T^{(r)}(z)T^{(s)}(w) &= 
    \frac{\delta_{r+s,0}\, Nc/2}{(z-w)^4} + \frac{2T^{(r+s)}(w)}{(z-w)^2} + \frac{\partial T^{(r+s)}(w)}{z-w}
    + \mathrm{reg}_{z \to w} \,, \\
    \Tb^{(r)}(\zb)\Tb^{(s)}(\wb) &= 
    \frac{\delta_{r+s,0}\, Nc/2}{(\zb-\wb)^4} + \frac{2\Tb^{(r+s)}(\wb)}{(\zb-\wb)^2} + \frac{\partial \Tb^{(r+s)}(\wb)}{\zb-\wb}
    + \mathrm{reg}_{\zb \to \wb} \,. 
  \end{aligned}
\end{equation}
The invariant modes $T^{(0)}(z)$ and $\Tb^{(0)}(\zb)$ are the components of the total stress-energy tensor of $\M_N$, with central charge $Nc$, whereas the other Fourier modes $T^{(r)}(z),\Tb^{(r)}(\zb)$ with $r \neq 0 \mod N$ can be regarded as additional conserved currents. Altogether, these Fourier modes encode an extended conformal symmetry.

The Laurent expansion of the currents $T^{(r)}(z)$ depends on the twist sector in which they act. Indeed, since by definition an operator $\Oc^{[k]}$ with charge $[k]$ inserts a defect line connecting copies $a$ and $a+k$, we have the monodromy conditions (in the absence of other twist operators in the vicinity of the origin)
\begin{equation}
\label{eq_semi_local_T}
  T^{(r)}(e^{2i\pi} z) \cdot \Oc^{[k]}(0) = \omega^{kr} T^{(r)}(z) \cdot \Oc^{[k]}(0)\,.
\end{equation}
As a result, the mode decomposition is of the form
\begin{equation} \label{eq:monodromy-T(r)}
   T^{(r)}(z) \cdot \Oc^{[k]}(0) = \sum_{m \in -kr/N+\Zbb} z^{-m-2} \, (L_m^{(r)} \cdot \Oc^{[k]})(0) \,\,.
  \end{equation}
  with $L_{n}^{(r) \dag} = L_{-n}^{(-r)}$. 
Likewise, for $\Tb^{(r)}$ we have
  \begin{equation}
 \Tb^{(r)}(\zb) \cdot \Oc^{[k]}(0) = \sum_{m \in +kr/N+\Zbb} \zb^{-m-2} \, (\Lb_m^{(r)} \cdot \Oc^{[k]})(0) \, \,.
\end{equation}
We stress that, in the twist sector $[k]$, the allowed indices for $L_m^{(r)}$ (resp. $\Lb_m^{(r)}$) are $m \in -kr/N+\Zbb$ (resp. $m \in +kr/N+\Zbb$).
The OPEs \eqref{eq:T(r).T(s)} yield the commutation relations
\begin{equation} \label{eq:OVir}
  \begin{aligned}
    & \left[ L^{(r)}_m, L^{(s)}_n \right] = (m-n) L_{m+n}^{(r+s)}
    + \frac{Nc}{12} m(m^2-1) \, \delta_{m+n,0} \, \delta_{r+s,0} \,, \\
    & \left[ \Lb^{(r)}_m, \Lb^{(s)}_n \right] = (m-n) \Lb_{m+n}^{(r+s)}
    + \frac{Nc}{12} m(m^2-1) \, \delta_{m+n,0} \, \delta_{r+s,0} \,, \\
    & \left[ L^{(r)}_m, \Lb^{(s)}_n \right] = 0 \,,
  \end{aligned}
\end{equation}
where the Kronecker symbols $\delta_{r+s,0}$ are understood modulo $N$.
These relations define the two commuting orbifold Virasoro algebras $\mathrm{OVir}_N$ and $\overline{\rm OVir}_N$.

The invariant modes $L_m^{(0)}$ are the modes of the total stress-energy tensor, and their index $m$ is always an integer, as it should be. Of course, they form a Virasoro algebra, with central charge $Nc$:
\begin{equation}
  \left[ L^{(0)}_m, L^{(0)}_n \right] = (m-n) L_{m+n}^{(0)}
    + \frac{Nc}{12} m(m^2-1) \, \delta_{m+n,0} \,, \qquad m \in \Zbb \,,
\end{equation}
and similarly for the $\Lb_m^{(0)}$'s.

Due to the conservation of Fourier indices in the commutation relations~\eqref{eq:OVir}, the two families of algebra elements
\begin{equation}
  L_{m_1}^{(r_1)} \dots L_{m_p}^{(r_p)} \quad \text{and} \quad \Lb_{m_1}^{(r_1)} \dots \Lb_{m_p}^{(r_p)} \,,
  \qquad \text{with }
    r_1+\dots+r_p = 0 \mod N 
\end{equation}
generate algebras of the universal enveloping algebra of $\mathrm{OVir}_N$ and $\overline{\rm OVir}_N$, which we shall call the \emph{neutral algebras} $A_N$ and $\Ab_N$, respectively. Recall that indices are restricted appropriately so that for the elements of the $A_N$ algebra we have that $m_i \in -kr_i/N+\Zbb$, and for $\Ab_N$ $m_i \in kr_i/N+\Zbb$. These neutral algebras are of key importance for the classification of operators and the description of correlations in the cyclic orbifold CFT. 

\subsection{Orbifold induction procedure}
\label{sec:induction}

As mentioned above, the Hilbert space $\mathcal{H}^{[1]}$ of the twisted sector $[k=1]$ is in one-to-one correspondence with the Hilbert space $\mathcal{H}$ of the mother theory \cite{borisov_systematic_1998}. More precisely, there is an isomorphism $\Theta_1 : \mathcal{H}  \to \mathcal{H}^{[1]}$, that is a norm-preserving, invertible linear map.
This map is in fact quite simple : it encodes the identification of states between $\mathcal{H}^{[1]}$ and  $\mathcal{H}$ that follows from unfolding the $N$ copies of the cylinder with twisted boundary-conditions into a single copy of the cylinder.  

In particular, we get the following identification between the stress-tensor $T$ of the mother theory (on the cylinder of perimeter $NL$) and the stress-tensor $T_a$ acting on the $a^{\textrm{th}}$ copy of the orbifold on the cylinder of perimeter $L$:
\begin{equation}
T_a(x,t) = \Theta_1 \, T( a L + x , t) \,\Theta_1^{-1} \,,
\end{equation}
where $0 \leq x < L$ and $a = 0, \cdots, N - 1$. In terms of modes, this means
\begin{equation}
  \left( L^{(r)}_m - \delta_{m,0} \frac{Nc}{24} \right)
  = \frac{1}{N}  \Theta_1 \left( L_{Nm} - \delta_{m,0} \frac{c}{24} \right) \Theta_1^{-1} \,,
  \qquad m \in -r/N+\Zbb \,,
\end{equation}
that is to say
\begin{equation}
  L^{(r)}_m =  \frac{1}{N}  \Theta_1 L_{Nm} \Theta_1^{-1} + \frac{c}{24} \left( N - \frac{1}{N}\right) \delta_{m0} \,,
  \qquad m \in -r/N+\Zbb \,.
\end{equation}
It follows that there is a one-to-one correspondence between Virasoro primary states $\ket{\phi_j}$ in the mother theory and primary states under the orbifold algebra in the twisted sector $[k=1]$ of the cyclic orbifold, which we will denote by $\ket{\sigma^{[1]}_j}$ :
\begin{equation}
\ket{\sigma^{[1]}_j} = \Theta_1 \ket{\phi_j} \,.
\end{equation}
The same elementary argument of finite-size scaling used to derive \eqref{eq:twist_dimension}
yields
\begin{equation}
\label{eq_dimension_composite_twist}
  h_{\sigma_j} =  \frac{c}{24} \left(N-\frac{1}{N} \right)  + \frac{h_j}{N} =   h_\sigma + \frac{h_j}{N} \,.
\end{equation}
The identification map $\Theta_1$ provides a comprehensive and explicit construction of the twisted sector $[k=1]$ in terms of states of the mother theory, known as the \emph{orbifold induction procedure} \cite{borisov_systematic_1998}. Furthermore, since $\Theta_1$ is norm-preserving, a state in the twisted sector is a null-state if and only if it is the image (under $\Theta_1$) of a null-state in the mother theory. In that sense, the induction procedure provides a full description of all null-states in the twisted sector. Such null-states are important, as they can be exploited to derive differential equations for twist correlation functions \cite{dupic_entanglement_2018}. 

Of course, the above discussion can be adapted to all twisted sectors $[k]$. In particular, when $k$ and $N$ are coprime (that is when the permutation $a \to a+k$ mod $N$ has a unique cycle), the Hilbert space $\mathcal{H}^{[k]}$ can still be identified with $\mathcal{H}$, via a map $\Theta_k : \mathcal{H} \to \mathcal{H}^{[k]}$. The relation between modes becomes simply
\begin{equation}
  L^{(r)}_m =  \frac{1}{N}  \Theta_k L_{Nm} \Theta_k^{-1}   + \frac{c}{24} \left( N - \frac{1}{N}\right) \delta_{m0} \,, \qquad m \in -kr/N+\Zbb \,.
\end{equation}
The state $\ket{\sigma^{[k]}_j} = \Theta_k \ket{\phi_j}$  has the same conformal dimension as $\ket{\sigma^{[1]}_j}$.

\section{Operator content, OPEs and conformal blocks}
\label{sec:op-algebra}

From now on, we restrict to the case when the number of copies is a prime integer $N$. We shall describe the operator content of the cyclic orbifold CFT $\M_N$, in terms of the primary operators $\{\phi_j\}$ of the mother CFT $\M$. In $\M_N$, operators and states will be organized into representations of the neutral algebras $A_N \oplus \Ab_N$ defined in Sec.~\ref{sec:OVir}. 

To simplify the discussion, we suppose that the mother CFT is diagonal, namely every primary operator $\phi_j$ is scalar, so that its conformal dimensions obey $h_j=\hb_j$, and all primaries correspond to distinct Virasoro modules. 
This is the case in particular for all diagonal minimal models.  Under these assumptions, the torus partition function is a diagonal modular invariant 
 \begin{equation}
Z = \sum_j \left| \chi_j \right|^2
\end{equation}
and the modular $S$-matrix is real symmetric. The fusion numbers $N_{ij}^k$, as given by the Verlinde formula
\begin{equation}
N_{i j }^k = \sum_m \frac{S_{im} S_{jm} \bar{S}_{mk}}{S_{\id m}}\,,
\end{equation}
 can only take the values $0$ or $1$.

\subsection{Invariant operators}
\label{sec:symm-operators}

Local operators in the $\mathbb{Z}_N$ orbifold must form a set of mutually local fields. In particular, they must be local with respect to the twist fields, \emph{i.e.} be invariant under cyclic permutations.  In that sense, gauging the ${\displaystyle \mathbb{Z}_N}$ symmetry in the replicated theory ${\displaystyle \mathcal{M}^{\otimes N}}$ removes \footnote{Such fields are not really removed : they are downgraded to semilocal fields in the cyclic orbifold. At the level of states, while they do not contribute to the torus partition function, they do contribute to twisted partition functions \cite{Petkova:2000ip}.} all fields in ${\displaystyle \mathcal{M}^{\otimes N}}$ that are not $\mathbb{Z}_N$ invariant. It turns out that this condition is also sufficient : all invariant fields are local, as we will show in section \ref{sec:torus_partition_function}. Throughout the paper, we will use the terms \emph{local} and \emph{invariant} interchangeably when referring to fields/operators.

Importantly, the currents $T^{(r)}(z)$  themselves are \emph{not local} for $r \neq 0$, since they have non-trivial monodromies around twist fields \eqref{eq_semi_local_T}.  This means that acting with a mode ${\displaystyle L^{(r)}_{m}}$ on a local field yields a non-local one, and therefore it is not possible to decompose local fields into modules of the full orbifold algebra $\mathrm{OVir}_N \oplus \overline{\rm OVir}_N$. Instead, one has to work with the neutral algebra $A_N$. 

Invariant operators can be classified into three families of primary operators and their descendants, with respect to the neutral algebras $A_N \oplus \Ab_N$. In this section, we enumerate all $(P^N-P)/N+ N^2 P$ primary invariant operators, where $P$ is the number of primary fields in the mother theory.

 The proofs that these operators are primary, and that the action of $A_N \oplus \Ab_N$ on them generates all invariant operators, are given in Appendix~\ref{sec:proof-primary-op}.

\paragraph{The non-diagonal untwisted operators $\Phi_{[j_1 \dots j_N]}$.}
They are defined as
\begin{equation} \label{eq:Phi-nondiag}
\Phi_J  =  \Phi_{[j_1,\cdots,j_N]} := \frac{1}{\sqrt N} \sum_{a=0}^{N-1} (\phi_{j_{1+a}} \otimes \dots \otimes \phi_{j_{N+a}}) \,, 
\end{equation}
where $J = [j_1,\cdots,j_N]$ stands for the $N$-tuple $(j_1,\cdots,j_N)$ modulo $\mathbb{Z}_N$, that is the equivalence class of $(j_1, \dots, j_N)$ under cyclic permutations. The indices $(1+a),\dots,(N+a)$ in \eqref{eq:Phi-nondiag} are understood modulo $N$.  Part of the definition is to demand that at least two labels among $(j_1,\cdots,j_N)$ are distinct, ensuring that the fields $\phi_{j_{1+a}} \otimes \dots \otimes \phi_{j_{N+a}}$, for $0 \leq a \leq N -1$ are linearly independent. Each $\phi_{j}$ stands for a primary operator in the mother CFT.  The conformal dimension of $\Phi_{J}$ is
\begin{equation}
  h_{[j_1,\cdots,j_N]} = h_{j_1} + \dots + h_{j_N} \,,
\end{equation}
where  $h_j$ is the conformal dimension of $\phi_j$. 

\paragraph{The diagonal untwisted operators $\Phi_j^{(r)}$, with $r \in \Zbb_N$.}
For $r=0$, let
\begin{equation}
  \Phi_j^{(0)} = \Phi_j := \phi_j \otimes \dots \otimes \phi_j \,,
\end{equation}
where $\phi_j$ is a primary operator in the mother CFT. For $r \neq 0 \mod N$, we define
\begin{align}
  & \Phi_j^{(r)} := \frac{1}{2Nh_j} L_{-1}^{(r)} \Lb_{-1}^{(-r)} \cdot \Phi_j
  \qquad \text{for } \phi_j \neq \id \,,  \label{eq:Phi_j} 
  \end{align}
and 
 \begin{align} 
  & \id^{(r)} :=  \frac{2}{Nc} L_{-2}^{(r)} \Lb_{-2}^{(-r)} \cdot \id  =  \frac{2}{Nc} T^{(r)} \overline{T}^{(-r)} \,.  \label{eq:Phi_1}
\end{align}
Recall that by convention, $\phi_1=\id$ in the mother CFT.
The reason for introducing a specific definition of the operators $\id^{(r)}$ is the fact that $L_{-1}^{(r)} \cdot \id =\Lb_{-1}^{(r)} \cdot \id = 0$, due to the null-vector conditions $L_{-1} \cdot \id=\Lb_{-1} \cdot \id=0$ in the mother CFT.
The corresponding conformal dimensions are
\begin{align}
  & h_j^{(r)} = Nh_j+ (1-\delta_{r0}) \qquad \text{for } \phi_j \neq \id \,, \\
  & h_1^{(r)} = 2(1-\delta_{r0}) \,.
\end{align}
The prefactors in (\ref{eq:Phi_j}--\ref{eq:Phi_1}) are chosen to normalize the two-point function – see below.

\paragraph{The twist operators $\sigma_j^{[k](r)}$.}

These operators are indexed by three labels : $j$ identifies a primary field $\phi_j$ in the mother theory,  $k \neq 0$ mod $N$ labels the twisted sector, and $r$ takes values in  $\Zbb_N$.  
When $j=1$ and $r=0$, they simply correspond to the ``bare'' twist operators
\begin{equation}
  \sigma_{\id}^{[k](0)} := \sigma^{[k]} \,,
\end{equation}
and they all have conformal dimension
\begin{equation}
  h_{\sigma} = \frac{c}{24} \left(N-\frac{1}{N} \right) \,.
\end{equation}
as follows from the state-operator correspondence and the finite size scaling argument above \eqref{eq:twist_dimension}. 
For $j \neq 1$ and $r=0$, the twist operator $  \sigma_{j}^{[k](0)} := \sigma_j^{[k]} $ is the field associated to the state {\small $\ket{\sigma^{[k]}_j}$} (as defined in section \eqref{sec:induction}) under the state-operator correspondence. This defines a \emph{composite twist operator}, with conformal dimension given by \eqref{eq_dimension_composite_twist}, that is :
\begin{equation}
  h_{\sigma_j} = h_\sigma + \frac{h_j}{N} \,.
\end{equation}
Alternatively, this composite twist field can be constructed as the most relevant field obtained in the fusion
\begin{equation}
 \Phi_{[j,\id,\dots,\id]} \times   \sigma_{\id}^{[k](0)}  
\end{equation}
where the field $ \Phi_{[j,\id,\dots,\id]}$ is defined at \eqref{eq:Phi-nondiag}. Formally \cite{Castro-Alvaredo_2011,Bianchini_2015} :
\begin{equation}\label{eq:compositetwistbuild}
  \sigma_j^{[k](0)}(z,\zb) =  \mathcal{A}_j \, \lim_{\epsilon \to 0} \left[ \epsilon^{2(1-N^{-1})h_j}
    \Phi_{[j,\id,\dots,\id]}(z+\epsilon,\zb+\bar\epsilon) \cdot \sigma^{[k]}(z,\zb)
    \right] \,,
\end{equation}
where $\Phi_{[j,\id,\dots,\id]}$ is the non-diagonal untwisted operator given above, and $h_j$ is the conformal dimension of $\phi_j$, and the constant prefactor is $\mathcal{A}_j = N^{-2(1-N^{-1})h_j-1/2}$. As explained in section \eqref{sec:induction}, the operators $\sigma_j^{[k]}$ are also primary under $\OVir_N \oplus \overline\OVir_N$.

For $r \neq 0$ the twist operators $\sigma_j^{[k](r)}$ are given by
\begin{equation}
  \sigma_j^{[k](r)} := \begin{cases}
    \mathcal{B}_{j,\br{kr}} \, L_{-\br{kr}/N}^{(r)} \Lb_{-\br{kr}/N}^{(-r)} \cdot \sigma_j^{[k]}
    & \text{if } \phi_j \neq \id \text{ or } kr \neq 1 \mod N \,, \\
    \mathcal{B}_{1,N+1} \, L_{-1-1/N}^{(r)} \Lb_{-1-1/N}^{(-r)} \cdot \sigma^{[k]}
    & \text{otherwise,}
  \end{cases}
\end{equation}
or equivalently as states 
\begin{equation}
  \ket{\sigma_j^{[k](r)}} := N^{-2} \Theta_k  \begin{cases}
    \mathcal{B}_{j,\br{kr}}  \, L_{-\br{kr}} \Lb_{-\br{kr}} \ket{\phi_j} 
    & \text{if } \phi_j \neq \id \text{ or } kr \neq 1 \mod N \,, \\
    \mathcal{B}_{1,N+1} \, L_{-(N+1)} \Lb_{-(N+1)} \ket{0}
    & \text{otherwise.}
  \end{cases}
\end{equation}
where $\br{n}$ stands for the remainder of the Euclidean division of $n$ by $N$, that is the unique integer in the interval $\{0,\dots,N-1\}$ such that $\br{n}=n \mod N$. The constant prefactors $\mathcal{B}_{j,n} = N^2\left[ 2 n h_j + c n (n^2-1)/12 \right]^{-1}$ are included to normalize the two-point function in the usual way \eqref{eq_normalization_2_pt_fn}.

The case $j = \id$ has to be treated separately because of the null-vector relations obeyed by the bare twist operators \cite{borisov_systematic_1998,dupic_entanglement_2018}:
\begin{equation}
  L_{-1/N}^{(k^{-1})} \cdot \sigma^{[k]} = 0 \,,
  \qquad \Lb_{-1/N}^{(-k^{-1})} \cdot \sigma^{[k]} = 0 \,.
\end{equation}
The conformal dimension of $\sigma_j^{[k](r)}$ is
\begin{equation}
  h_j^{[k](r)} = \begin{cases}
    h_{\sigma_j} + \frac{\br{kr}}{N}
    & \text{if } \phi_j \neq \id \text{ or } kr \neq 1 \mod N \,, \\
    h_{\sigma} + \frac{N+1}{N}
    & \text{otherwise.}
  \end{cases}
\end{equation}

\paragraph{Property.} All the operators in the above list are primary under the neutral algebra $A_N \oplus \Ab_N$. They are local, by virtue of being invariant under cyclic permutations of copies. They are scalar operators, namely they have conformal dimensions $h=\hb$. Moreover, there are the only fields with these properties : any invariant operator in $\M_N$ can be obtained by acting with $A_N \oplus \Ab_N$ on one of these invariant primary operators. These fundamental properties are proven in Appendix~\ref{sec:proof-primary-op}.

\paragraph{Two-point functions.} The normalization factors have been chosen so that the two-point functions are given by
\begin{equation}
  \begin{aligned}
    &\aver{\Phi_{[j_1,\dots,j_N]}(z,\zb)\Phi_{[j_1,\dots,j_N]}(w,\wb)}
    = |z-w|^{-4h_{[j_1,\dots,j_N]}} \,, \\
    &\aver{\Phi_j^{(r)}(z,\zb)\Phi_j^{(-r)}(w,\wb)}
    = |z-w|^{-4h_j^{(r)}} \,, \\
    & \aver{\sigma_j^{[k](r)}(z,\zb)\sigma_j^{[-k](-r)}(w,\wb)}
    = |z-w|^{-4h_j^{[k](r)}} \,,
  \end{aligned}
\end{equation}
where $r \in \Zbb_N$ and $k \in \Zbb_N^\times$. Any other two-point function of invariant primary operators vanishes. This can be summarized by writing
\begin{equation}
\label{eq_normalization_2_pt_fn}
  \aver{\Phi_\alpha(z,\zb) \Phi_\beta^\dag(w,\wb)} = \delta_{\alpha\beta} \, |z-w|^{-4h_\alpha} \,,
\end{equation}
where $\Phi_\alpha$ denotes an operator of the form $\Phi_{[j_1 \dots j_N]}$, $\Phi_j^{(r)}$ or $\sigma_j^{[k](r)}$, and the conjugation $\alpha \to \alpha^\dag$ acts as
\begin{align}
  [j_1 \dots j_N] & \to [j_1 \dots j_N]  \\
   j ,(r) & \to  j, (-r) \\
   j,[k],(r) & \to j,[-k],(-r) 
\end{align}
\textbf{Counter example to classification for non-prime $N$.}  The classification of primary operators mentioned earlier is valid only for prime values of $N$. To demonstrate this, we examine the case of $N=4$ in the non-diagonal sector and present a primary field that defies the classification discussed above. Consider the invariant field 
\begin{equation}
\Phi_{[1,2,1,2]} = (\phi_1 \otimes 
\phi_2 \otimes\phi_1 \otimes\phi_2 + \phi_2 \otimes \phi_1 \otimes\phi_2 
\otimes\phi_1)
\end{equation}
for which one can check that:
\begin{equation}
L_0^{(1)} \cdot \Phi_{[1,2,1,2]} = L_0^{(3)} \cdot \Phi_{[1,2,1,2]} = 0
\end{equation}
The equivalent relation for antiholomorphic modes also holds.
By adapting the arguments of Appendix \ref{app:untwistedsector_primary} for non-diagonal fields, one can then show that
\begin{equation}
L_{-1}^{(1)} \bar{L}_{-1}^{(3)} \cdot \Phi_{[1,2,1,2]}
\end{equation}
is also primary under the neutral algebra.

\subsection{Invariant Hilbert space}
\label{sec:Hilbert}

In any \emph{chiral} $\OVir_N$-module with a lowest-weight state $\ket{h_\alpha}$, we can define a formal cyclic permutation operator $\pi$ by
\begin{equation}
  \pi \cdot \ket{h_\alpha} := \ket{h_\alpha} \,,
  \qquad \pi \cdot L_m^{(r)} := \omega^r \, L_m^{(r)} \cdot\pi \,,
\end{equation}
and similarly for an $\overline{\OVir}_N$-module.
With this definition, we have $\pi^N=\id$, and hence the operator
\begin{equation}
  P_r = \frac{1}{N} \sum_{a=0}^{N-1} \omega^{-ar} \, \pi^a \,.
\end{equation}
is the projector on the eigenspace of $\pi$ corresponding to the eigenvalue $\omega^r$. 

For any invariant primary operator $\Phi_\alpha$ of the form $\Phi_{[j_1 \dots j_N]}$, $\Phi_j^{(r)}$ or $\sigma_j^{[k](r)}$, we denote by $\Vc_\alpha$ the corresponding $(A_N \oplus \Ab_N)$-module. The action of $A_N$ (resp. $\Ab_N$)  on the highest weight state $\ket{h_\alpha}$ generates an $A_N$-module (resp. $\Ab_N$-module), which we denote as $V_\alpha$ (resp. $\Vb_\alpha$). Since $A_N$ and $\Ab_N$ commute, we have
\begin{equation}
  \Vc_\alpha \simeq V_\alpha \otimes \Vb_\alpha \,.
\end{equation}
It turns out that the $A_N$-modules $V_\alpha$ associated to the three types of invariant primary operators $\Phi_\alpha$ in the orbifold model can be viewed as eigenspaces of $\pi$ in $\OVir_N$-modules, namely
\begin{equation}
  V_{[j_1,\dots,j_N]} \simeq  P_0 \cdot (v_{j_1} \otimes \dots \otimes v_{j_N}) \,,
  \quad V_j^{(r)} \simeq  P_r \cdot (v_j \otimes \dots \otimes v_j) \,,
  \quad V_j^{[k](r)} \simeq  P_r \cdot \Theta_k \cdot v_j \,.
\end{equation}
Here, we have denoted by $v_j$ the $\Vir$-module associated to the primary operator $\phi_j$ in the mother CFT, and we have used the induction isomorphisms $\Theta_k$ introduced in Sec.~\ref{sec:induction}. For $\Ab_N$-modules we have
\begin{equation}
  \Vb_{[j_1,\dots,j_N]} \simeq  P_0 \cdot (\vb_{j_1} \otimes \dots \otimes \vb_{j_N}) \,,
  \quad \Vb_j^{(r)} \simeq  P_{-r} \cdot (\vb_j \otimes \dots \otimes \vb_j) \,,
  \quad \Vb_j^{[k](r)} \simeq  P_{-r} \cdot \Theta_k \cdot \vb_j \,,
\end{equation}
where the $\vb_j$'s are the mother CFT's $\overline{\Vir}$-modules.

Let $\mathcal{H}_0$ be the space of $\M_N$ states which are invariant under cyclic permutations.
We shall refer to $\mathcal{H}_0$ as the \emph{invariant Hilbert space}.
As a consequence of the classification of invariant operators presented in Sec.~\ref{sec:symm-operators}, one gets the $(A_N \oplus \Ab_N)$-module decomposition of $\mathcal{H}_0$:
\begin{equation} \label{eq:H0}
  \mathcal{H}_0 = \bigoplus_{J = [j_1 \dots j_N]} \Vc_{J}
  \quad\oplus\quad \bigoplus_j \bigoplus_{r=0}^{N-1} \Vc_j^{(r)}
  \quad\oplus\quad \bigoplus_j \bigoplus_{k=1}^{N-1} \bigoplus_{r=0}^{N-1} \Vc_j^{[k](r)} \,.
\end{equation}
In this expression, the sums on $j$ run over the primary operators $\phi_j$ of the mother CFT, whereas the sum on $[j_1 \dots j_N]$ runs over the equivalence classes of $N$-tuples $(j_1, \dots, j_N)$ under cyclic permutations, where the $j_a$'s are indices of primary operators of the mother CFT, and with at least two distinct indices $j_a \neq j_b$.

\subsection{On fusion numbers}
\label{sec:correlation-functions}

Before moving on to operator product expansions and conformal blocks, it is useful to discuss fusion numbers. From the above discussion, the cyclic orbifold is a diagonal theory with respect to the extended symmetry $A_N \oplus \Ab_N$, therefore operators $\{\Phi_{\alpha}\}$ are labelled by a single symbol $\alpha$, also labelling the irreducible $A_N$-modules $\{V_\alpha\}$. The fusion rules are generically of the form 
\begin{equation}
  \Phi_\alpha \times \Phi_\beta \to \sum_\gamma \Nc_{\alpha\beta}^\gamma  \, \Phi_\gamma \,,
\end{equation}
where the sum is over all primary invariant operators $\Phi_\gamma$, and the non-negative integers $\Nc_{\alpha \beta}^\gamma$ are the fusion numbers. They obey 
\begin{equation}
  \Nc_{\alpha\beta}^\gamma = \Nc_{\beta\alpha}^\gamma \,,
  \qquad \Nc_{\alpha\beta}^\gamma = \Nc_{\alpha\gamma^\dag}^{\beta^\dag} \,.
\end{equation}
Recall that we are assuming that, in the mother theory, all fusion numbers $N_{ij}^k$ are $0$ or $1$. But in the orbifold, non-trivial multiplicities (\emph{i.e.} $\Nc_{\alpha \beta}^\gamma > 1$) are expected to appear. The fusion number $\mathcal{N}_{\alpha\beta}^\gamma$ is the dimension of the space of chiral vertex operators \cite{1988PhLB..212..451M} of type {\tiny $\left( {}_{\alpha\beta}^{ \, \,  \gamma} \right)$}. Chiral three point functions 
\begin{equation}
  \mathcal{C}_{\alpha\beta}^\gamma(\lambda,\mu,\nu) = \aver{\Phi_\gamma|\nu^\dag  \, (\lambda \cdot \Phi_\alpha)(1) \,  \mu  |\Phi_\beta}
\end{equation}
for $\lambda, \mu ,\nu \in A_N$ are not all linearly independent. Indeed, they satisfy many linear relations following from:
\begin{enumerate}
\item The commutation rules of $A_N$, which follow from
  \begin{equation} \label{eq:[Lm,Ln]}
    \left[ L^{(r)}_m, L^{(s)}_n \right] = (m-n) L_{m+n}^{(r+s)}
    + \frac{Nc}{12} m(m^2-1) \, \delta_{m+n,0} \, \delta_{r+s,0} \,.
  \end{equation}
\item The properties of primary operators and states
  \begin{align}
    & [L_m^{(0)},\Phi_\alpha(z,\zb)] = z^m \, \Big((m+1)h_\alpha + z\partial_z \Big) \, \Phi_\alpha(z,\zb) \,,
      \label{eq:[Lm,Phi]} \\
    & \bra{\Phi_\gamma} L_{m<0}^{(r)} = 0 \,, \qquad \bra{\Phi_\gamma} L_0^{(0)} = h_\gamma \bra{\Phi_\gamma} \,,
    \label{eq:<phi|Lm} \\
    & L_{m>0}^{(r)}\ket{\Phi_\alpha}=0 \,, \qquad  L_0^{(0)}\ket{\Phi_\alpha} = h_\alpha \ket{\Phi_\alpha} \,.
      \label{eq:Lm|phi>} 
  \end{align}
\item The Ward identities associated to the $\OVir_N$ currents.
  These can be expressed generally as  closed contour identities, for any $n$-tuple of operators $\Phi_1,\dots,\Phi_n$ (primary or not) with twist charges $k_1,\dots,k_n$:
  \begin{align}
    &\oint dz \, (z-z_1)^{q_1}\dots (z-z_n)^{q_n}
    \, \aver{T^{(r)}(z) \Phi_1(z_1,\zb_1)\dots\Phi_n(z_n,\zb_n)} = 0 \,, \nn \\
    &\qquad\qquad \text{if } q_j \in -\frac{rk_j}{N}+\Zbb \text{ and } q_1+\dots+q_n \leq 2 \,.
      \label{eq:orbifold-Ward}
  \end{align}
  \item The decoupling of null-vectors (\emph{e.g.} if $ \mu  |\Phi_\beta \rangle $ is null, then  $  \mathcal{C}_{\alpha\beta}^\gamma(\lambda,\mu,\nu) =0$). 
\end{enumerate}
For instance, using \eqref{eq:[Lm,Phi]} and \eqref{eq:Lm|phi>}, we have for $m>0$:
\begin{equation}
  \mathcal{C}_{\alpha\beta}^\gamma(\id,\id,L_{-m}^{(0)})
  = \aver{\Phi_\gamma|L_m \Phi_\alpha(1) |\Phi_\beta}
  = \Big((m+1)h_\alpha -2h_\alpha-2h_\beta+2h_\gamma \Big) \, \mathcal{C}_{\alpha\beta}^\gamma(\id,\id,\id) \,.
\end{equation}
Consider the whole set of constraints (\ref{eq:[Lm,Ln]}--\ref{eq:orbifold-Ward}) plus the decoupling of null-vectors as a linear system of equations for the function
$$
\begin{cases}
  A_N \otimes A_N \otimes A_N &\to \quad \Cbb \\
  (\lambda,\mu,\nu) &\mapsto \quad  \mathcal{C}_{\alpha\beta}^\gamma(\lambda,\mu,\nu) =  \aver{\Phi_\gamma|\nu^\dag  \, (\lambda \cdot \Phi_\alpha)(1) \,  \mu  |\Phi_\beta} \,.
\end{cases}
$$
Then $\mathcal{N}_{\alpha\beta}^\gamma$ is the dimension of the solution space of this linear system.
By the use of orbifold Ward identities, one can show that the subset of coefficients
\begin{equation}
  \mathcal{C}_{\alpha\beta}^\gamma(\mu):= \mathcal{C}_{\alpha\beta}^\gamma(\id,\mu,\id)
  = \aver{\Phi_\gamma| \Phi_\alpha(1) \mu |\Phi_\beta} \,,
\end{equation}
determines uniquely all the other coefficients $\mathcal{C}_{\alpha\beta}^\gamma(\lambda,\mu,\nu)$. Hence,
$\mathcal{N}_{\alpha\beta}^\gamma$ corresponds to the number of linearly independent solutions for the function $\mu \mapsto \mathcal{C}_{\alpha\beta}^\gamma(\mu)$, subject to (\ref{eq:[Lm,Ln]}--\ref{eq:orbifold-Ward}). Let $\left( \mathcal{C}_{\alpha\beta} \right)_m$, with $m =1 , \cdots, \mathcal{N}_{\alpha\beta}^\gamma$, denote a basis of these solutions.
\bigskip

If we now return to the physical three point function (that is including the anti-holomorphic degrees of freedom), generically we have
\begin{equation}
  \aver{\Phi_\gamma| \Phi_\alpha(1,\bar{1}) \mu \bar{\mu} |\Phi_\beta}  = \sum_{m,n=1}^{ \mathcal{N}_{\alpha\beta}^\gamma}  \kappa_{m,n} \left( \mathcal{C}^{\gamma}_{\alpha\beta} \right)_m(\mu) \left( \mathcal{C}^{\gamma}_{\alpha\beta} \right)_n (\bar{\mu}) \,. 
\end{equation}
However since the application $(\mu , \bar{\mu}) \to  \aver{\Phi_\gamma| \Phi_\alpha(1,\bar{1}) \mu \bar{\mu} |\Phi_\beta}$ is bilinear and symmetric, there exists a (real) basis $\left( X_{\alpha\beta} \right)_m$, such that 
 \begin{equation}
 \label{eq:3pt-mu-mub}
  \aver{\Phi_\gamma| \Phi_\alpha(1,\bar{1}) \mu \bar{\mu} |\Phi_\beta}  = \sum_{m=1}^{ \mathcal{N}_{\alpha\beta}^\gamma}  \epsilon_m \left(X^{\gamma}_{\alpha\beta} \right)_m(\mu) \left( X^{\gamma}_{\alpha\beta} \right)_m (\bar{\mu})   \,,
\end{equation}
where $\epsilon_m \in \{ -1, 0 ,1 \}$. The naturality theorem \cite{MOORE198916} implies that the above bilinear form has maximal rank, thus $\epsilon_m =0$ must be excluded.

\bigskip
Let's consider for instance the fusion process involving three non-diagonal untwisted operators
\begin{equation} \label{eq:fusion[i][j][k]}
  \Phi_{[i_1\dots i_N]} \times \Phi_{[j_1\dots j_N]} \to \Phi_{[k_1\dots k_N]} \,.
\end{equation}
For compactness we introduce the shorthand notation $I = [i_1\dots i_N], J = [j_1\dots j_N]$ and $K = [k_1\dots k_N]$. 
Consider the following physical three point function 
\begin{equation}
  \mathcal{C}_{I J}^K(\mu,\bar{\mu})
  = \aver{\Phi_{K} \Phi_{I} (\mu \, \bar{\mu} \cdot\Phi_{J})} \,,
\end{equation}
for $\mu \in A_N$ and $\bar{\mu} \in \bar{A}_N$. Using the definition of $\Phi_{[i_1\dots i_N]},\Phi_{[j_1\dots j_N]},\Phi_{[k_1\dots k_N]}$ and the fact that $\mu, \bar{\mu}$ are invariant under the cyclic permutation $\Pi$, one can write
\begin{equation} \label{eq:C[i][j][k]}
  \mathcal{C}_{I J}^K(\mu,\bar{\mu})
  = \sum_{a,b=0}^{N-1} \left( \mathcal{C}_{I J}^K\right)_{ab}(\mu,\bar{\mu}) \,,
\end{equation}
where (with a slight abuse of notation, as we now treat $I,J$ and $K$ as $N$-tuples)
\begin{equation}
   \left( \mathcal{C}_{I J}^K\right)_{ab}(\mu,\bar{\mu}):=
  \frac{1}{\sqrt N} \aver{\Pi^a[\phi_{k_1} \otimes \dots\otimes\phi_{k_N}]
    \, \Pi^b[\phi_{i_1} \otimes \dots\otimes\phi_{i_N}] \, (\mu\, \bar{\mu}\cdot(\phi_{j_1} \otimes \dots\otimes\phi_{j_N}))} \,.
\end{equation}
and $\Pi$ is the generator of cyclic permutations\footnote{This is not to be confused with $\pi$, which acts in the \textit{chiral} $OVir_N$ module. The operator $\Pi$ acts on physical invariant primary operators.} :  $\Pi [\phi_{i_1} \otimes \dots\otimes\phi_{i_N}] = [\phi_{i_2} \otimes \phi_{i_3} \otimes \dots\otimes\phi_{i_1}]$. A closer inspection reveals that each symmetric bilinear form $\left( \mathcal{C}_{I J}^K\right)_{ab}$ is rank at most one. Indeed, decomposing
\begin{equation}
\mu =   L_{m_1}^{(r_1)} \dots L_{m_p}^{(r_p)} 
\end{equation}
together with 
\begin{equation}
L_m^{(r)} = \sum_{j=0}^{N-1} \omega^{j r} \,  \id \otimes \dots \otimes \underset{(j)}{L_m}\otimes \dots \otimes\id \,,
\end{equation}
and likewise for $\bar{\mu}$, one ends up with a sum of products of three point functions in the mother theory of the form
\begin{equation}
\prod_c \aver{\phi_{k_{c+a}}\phi_{i_{c+b}}  \, \left(   \lambda_c   \bar{\lambda}_c \cdot  \phi_{j_c} \right) }
\end{equation}
where $\lambda, \bar{\lambda}$ are in (the enveloping algebra of) Virasoro. For each term in the product, we have 
\begin{equation}
  \aver{\phi_k \phi_i (\lambda \, \bar{\lambda} \cdot \phi_j)}
  = \begin{cases}
    \aver{\phi_k \phi_i \phi_j} \, x_{ij}^k(\lambda)  x_{ij}^k(\bar{\lambda}) & \text{if } N_{ij}^k=1 \,, \\
    0 & \text{if } N_{ij}^k=0 \,,
  \end{cases}
\end{equation}
where $x_{ij}^k(\lambda)$ is uniquely determined by the Virasoro analogue of (\ref{eq:[Lm,Ln]}--\ref{eq:orbifold-Ward}). Clearly  $\left( \mathcal{C}_{I J}^K\right)_{ab}$ vanishes unless 
\begin{equation}
\label{eq:fusion_constraint}
 \aver{\phi_{k_{1+a}}\phi_{i_{1+b}}\phi_{j_1}} \dots \aver{\phi_{k_{c+a}}\phi_{i_{c+b}} \phi_{j_c} }  \dots \aver{\phi_{k_{N+a}}\phi_{i_{N+b}}\phi_{j_N}} \neq 0
\end{equation}
Moreover when it does not vanish, it factorizes as
\begin{equation}
   \left( \mathcal{C}_{I J}^K\right)_{ab}(\mu,\bar{\mu}):=        \left(X_{I J}^K\right)_{ab}(\mu)   \left(X_{I J}^K\right)_{ab}(\bar{\mu})
\end{equation}
where the linear forms $  \left(X_{I J}^K\right)_{ab}$ can be computed explicitly in terms of the  $x_{ij}^k$'s.
For instance, one has
\begin{align}
   \left( X_{I J}^K\right)_{ab}(L_m^{(0)}) \propto \sum_{c=0}^{N-1} x_{i_{c+b},j_c}^{k_{c+a}}(L_m)
\end{align}
and
\begin{align}
  &  \left( X_{I J}^K\right)_{ab}(L_m^{(r)}L_n^{(-r)}) \nn\\
  & \qquad \propto \sum_{c,d=0}^{N-1} \Big[ (1-\delta_{cd})\omega^{(c-d)r} x_{i_{c+b},j_c}^{k_{c+a}}(L_m)x_{i_{d+b},j_d}^{k_{d+a}}(L_n)
  + \delta_{cd} x_{i_{c+b},j_c}^{k_{c+a}}(L_mL_n) \Big] \,.
\end{align}

\bigskip

We have decomposed the physical three point function $  \aver{\Phi_\gamma| \Phi_\alpha(1,\bar{1}) \mu \bar{\mu} |\Phi_\beta} $ in terms of a family of solutions $ \left( X_{I J}^K\right)_{ab}(\mu)$ of (\ref{eq:[Lm,Ln]}--\ref{eq:orbifold-Ward}), and shown how to compute them. The cardinal of this family is
\begin{equation} \label{eq:N[i][j][k]}
 \sum_{a,b}  N_{i_{1+a},j_{1+b}}^{k_1} \dots N_{i_{N+a},j_{N+b}}^{k_N} \,,
\end{equation}
where $N_{ij}^k$ is the fusion number for $\phi_i \times \phi_j \to \phi_k$ in the mother CFT. By the naturality theorem \cite{MOORE198916}, the forms $ \left( X_{I J}^K\right)_{ab}(\mu)$ span the whole space of solutions. But, as we did not prove that the $ \left( X_{I J}^K\right)_{ab}(\mu)$ are linearly independent, \eqref{eq:N[i][j][k]} is only an upper bound for the fusion number $\mathcal{N}_{IJ}^K$. The fact that   $\mathcal{N}_{IJ}^K$ indeed equals \eqref{eq:N[i][j][k]}, and therefore the $ \left( X_{I J}^K\right)_{ab}(\mu)$ are linearly independent, will be proved indirectly in Sec.~\ref{sec:fusion-rules} via the Verlinde formula. 

 \bigskip

The above arguments can be easily extended to construct a family of independent OPE coefficients $(\mathcal{C}_{\alpha\beta}^\gamma)_m(\mu)$ for a fusion process involving any kind of untwisted operators, leading to the fusion numbers
\begin{align}
  & \Nc_{[i_1 \dots i_N],[j_1 \dots j_N]}^{[k_1\dots k_N]} =  \sum_{a,b}  N_{i_{1+a},j_{1+b}}^{k_1} \dots N_{i_{N+a},j_{N+b}}^{k_N} \,, \\
  & \Nc_{[i_1 \dots i_N],[j_1 \dots j_N]}^{k^{(r)}} = \sum_{a=0}^{N-1} N_{i_{1+a},j_1}^k \dots N_{i_{N+a},j_N}^k \,, \\
  & \Nc_{[i_1 \dots i_N],j^{(r)}}^{k^{(s)}} = N_{i_1,j}^k \dots N_{i_N,j}^k \,, \\
  & \Nc_{j^{(r)},k^{(s)}}^{\ell^{(t)}} = \delta_{r+s,t} \, N_{ij}^k \,.
\end{align}
Fusion number involving twist fields on the other hand cannot be inferred using the same elementary approach, because for such three point functions the copies of the mother CFT within the orbifold CFT are no longer decoupled. The answer, as provided by the Verlinde's formula, will turn out to be more complicated, and not expressible in terms of the fusion numbers of the mother theory alone. But while we cannot at this stage easily compute the fusion numbers involving twists fields, we can at least argue that they are finite, by showing that correlation functions in the orbifold always involve finitely many (extended) conformal blocks.

\paragraph{Holomorphic factorization property and rationality}

Conformal field theories obey the holomorphic factorization property \cite{friedan1987analytic}. On the plane/sphere, it means that  
correlation functions can be decomposed as 
\begin{equation} 
\langle \phi_{i_1} (z_1,\bar{z}_1) \cdots \phi_{i_n} (z_n,\bar{z}_n) \rangle = \sum_{I,J} F_I ( {\bf z}) \rho_{I J} \overline{F}_{J} ( {\bf \bar{z}}) 
\end{equation}
where the (extended) conformal blocks $F_I$ are holomorphic functions of  ${\bf z}  = (z_1,\cdots, z_n)$, and $\overline{F}_{J}$ are anti-holomorphic. More generally, on a Riemann surface $\Sigma$ the factorization property becomes 
   \begin{equation} 
Z_{\Sigma} \,  \,  \langle \phi_{i_1} (z_1,\bar{z}_1) \cdots \phi_{i_n} (z_n,\bar{z}_n) \rangle_{\Sigma} = \sum_{I,J} F_I ( {\bf z},{\bf m}) \rho_{I J} \overline{F}_{J} ( {\bf \bar{z}}, {\bf \bar{m}})  
\end{equation}
where $Z_{\Sigma}$ stands for the partition function on $\Sigma$, and the conformal blocks $F_I$ depend also holomorphically on the analytic coordinates ${\bf m}$ of the moduli of $\Sigma$. The above holds in particular in the absence of field insertions ($n=0$), meaning that the partition function itself obeys the holomorphic factorization property. For instance in genus one, with a flat metric, this is simply 
   \begin{equation} 
Z( \tau, \bar{\tau})  = \sum_{i,j} \chi_i(\tau)  \rho_{ij } \overline{\chi}_{j} (\bar{\tau}) \,.
\end{equation}
A CFT is said to be \emph{rational} when all the sums involved above are finite. 

Provided the mother theory $\mathcal{M}$ obeys the holomorphic factorization property, then so does its cyclic orbifold $\mathcal{M}_N$. Consider for instance a generic correlation function of twists fields on a Riemann surface $\Sigma$
   \begin{equation} 
 \langle \sigma^{[k_1]}_{i_1} (z_1,\bar{z}_1) \cdots \sigma^{[k_n]}_{i_n} (z_n,\bar{z}_n) \rangle_{\Sigma} 
\end{equation}
where twist charge neutrality is assumed ($\sum_i k_i =0$ mod $N$), and $\sigma^{[k]}_i$ stand for the composite twist field of charge $k$ associated to the field $\phi_i$ (as defined in section \ref{sec:symm-operators}).

One can reinterpret this correlation function as a correlation function of the fields $\phi_i$ in the mother theory on the $N$-sheeted branched cover $p :  \Sigma' \to \Sigma$, with appropriate branch points at $z_j$ : 
   \begin{align} 
Z_{\Sigma} \,  \,  \langle \sigma^{[k_1]}_{i_1} (z_1,\bar{z}_1) \cdots \sigma^{[k_n]}_{i_n} (z_n,\bar{z}_n) \rangle_{\Sigma}   & = Z_{\Sigma'} \langle  \phi_{i_1} (z_1,\bar{z}_1) \cdots \phi_{i_n} (z_n,\bar{z}_n) \rangle_{\Sigma'} \,,
\end{align}
where in the \emph{r.h.s.} both the correlation function and the partition function involve the mother theory on the covering surface $\Sigma'$, with the metric $g' = p^*g$ being the pull-back of the metric $g$ on $\Sigma$. In practice, the picture above is slightly more complicated, as the pull-back metric $g'$ has conical singularities at the ramification points, but these can be regularized \cite{Lunin:2000yv}.

Exploiting the holomorphic factorization property of the mother theory then yields 
   \begin{align} 
Z_{\Sigma} \,  \,  \langle \sigma^{k_1}_{i_1} (z_1,\bar{z}_1) \cdots \sigma^{k_n}_{i_n} (z_n,\bar{z}_n) \rangle_{\Sigma}  =  \sum_{I,J} F_I ( {\bf z},{\bf m}') \rho_{I J} \overline{F}_{J} ( {\bf \bar{z}}, {\bf \bar{m}'}) 
\end{align}
The factorization property of the \emph{l.h.s.} then follows from the fact that the moduli ${\bf m'} ={\bf m'}({\bf m}, {\bf z})$ of the branched covering $\Sigma'$ depends holomorphically on the moduli ${\bf m}$ of $\Sigma$ and the positions ${\bf z}$ of the branched points. For instance, the $2$-sheeted cover of the sphere with four branch points at positions $z_i$  is a torus with moduli
\begin{align*}
 \tau(x) = i\,  \frac{{}_2\mathrm{F}_{1} \left( \frac{1}{2} , \frac{1}{2} , 1 ; 1-  x \right)}
       {{}_2\mathrm{F}_{1} \left( \frac{1}{2} , \frac{1}{2} , 1 ; x \right)}, \qquad  x = \frac{(z_1-z_2)(z_3-z_4)}{(z_1-z_3)(z_2-z_4)}
\end{align*}

The above argument is easily adapted to the generic case involving descendants of twist fields and/or untwisted fields. The last quantity to consider is the partition function itself, but this is simply obtained by summing all  twisted partition functions of $N$ copies of the mother theory, each of which obeys the  factorization property.

Furthermore, it is clear that if the mother theory is rational, then so is the orbifold. As a consequence, all fusion numbers in the cyclic orbifold are finite when the mother theory is rational.

\subsection{(Extended) conformal blocks on the sphere}
\label{sec:conformal-blocks}
 
The computation of entanglement entropies in the case of a single interval embedded in an infinite line (or a circle) at zero temperature boils down to a two-point function of twist fields on the sphere \cite{holzhey_geometric_1994,calabrese_entanglement_2004,calabrese_entanglement_2009}. The main simplification in that case is that such two-point functions are completely fixed by conformal invariance. In contrast, if the system has boundaries, or is at finite temperature with periodic boundary conditions, the CFT computation involves two point functions of twist fields on more complicated Riemann surfaces, namely, the upper half plane or the torus\cite{Cardy:2014jwa,sully_bcft_2021,2022ScPP...12..141E,datta_renyi_2014,Chen2014SingleIR,2016JHEP...01..058L,mukhi_entanglement_2018,Wu2019FiniteTE,Gerbershagen2021MonodromyMF,Rotaru_compact_boson,Estienne:2023tdw}. Likewise, if the region $A$ consists of several disjoint intervals, Rényi entropies involve higher-point correlation functions of twist fields, or equivalently the partition function on higher genus Riemann surfaces \cite{2008JHEP...11..076C,Furukawa:2008uk,fagotti_entanglement_2010,calabrese_entanglement_2009-1,headrick_entanglement_2010,Calabrese:2010gpa,alba_entanglement_2010,Calabrese_2011,alba_entanglement_2011,Calabrese:2012ew,calabrese_entanglement_2013,coser_renyi_2014,2015JSMTE..06..021D,2016JSMTE..05.3109C,Coser:2015eba,Grava:2021yjp}. In such cases, conformal invariance alone is no longer sufficient, and the full resources of two-dimensional CFT have to be brought to bear on the computation. One practical approach is to expand correlation functions using standard Virasoro conformal blocks, either to produce asymptotic expansions in some limiting cases (such as a small interval, for instance) \cite{Calabrese_2011,rajabpour_entanglement_2012,2014JHEP...09..010K,Chen_2016,Li:2016pwu,2016JHEP...11..116L,Ruggiero_2018} or to implement a numerical bootstrap approach\cite{ares_crossing-symmetric_2021}. However, by building on the present work, one can exploit the extended $A_N$ symmetry and use extended conformal blocks instead of a brute force expansion employing just Virasoro blocks.

As we shall now explain, the diagonal decomposition~\eqref{eq:3pt-mu-mub} of OPE coefficients is a key property for the analysis of correlation functions, and allows for a systematic description of orbifold conformal blocks. 
Consider the four-point function
\begin{equation}
  G(z,\zb) = \aver{\Phi_1^\dag(\infty)\Phi_2^\nodag(1)\Phi_3^\nodag(z,\zb)\Phi_4^\nodag(0)} \,,
\end{equation}
where $\Phi_1, \dots, \Phi_4$ are (twisted or untwisted) invariant primary operators in $\M_N$.
Inserting a resolution of the identity, we get
\begin{align}
  G(z,\zb) = \sum_{\alpha,\mu,\mub} z^{-h_{34}^\alpha+|\mu|} \, z^{-h_{34}^\alpha+|\mub|}
  \, \aver{\Phi_1^\dag \Phi_2^\nodag (\mu \cdot\mub \cdot\Phi_\alpha)}
  \, \aver{(\mu \cdot\mub \cdot\Phi_\alpha)^\dag \Phi_3^\nodag \Phi_4^\nodag} \,,
\end{align}
where we have used the notation $h_{ab}^c= h_a+h_b-h_c$, and $|\mu|,|\mub|$ denote the levels of $\mu,\mub$ respectively.
In this expansion, $\Phi_\alpha$ runs over all the invariant primary operators, and $\{\mu\ket{\Phi_\alpha}\}$ (resp. $\{\mub\ket{\Phi_\alpha}\}$) is an \emph{orthonormal} basis of $V_\alpha$ (resp. $\Vb_{\alpha}$).
We now use the decomposition~\eqref{eq:3pt-mu-mub} for the three-point functions, which gives
\begin{align}
  & \aver{\Phi_1^\dag \Phi_2^\nodag (\mu \cdot\mub \cdot\Phi_\alpha)}
  = \sum_{m=1}^{\Nc_{2\alpha}^1} (\epsilon_{2\alpha}^1)_m
  \, \left( X^{1}_{2\alpha} \right)_m(\mu) \, \left( X^{1}_{2\alpha} \right)_m(\mub) \,, \\
  &\aver{(\mu \cdot\mub \cdot\Phi_\alpha)^\dag \Phi_3^\nodag \Phi_4^\nodag}
  = \sum_{n=1}^{\Nc_{34}^\alpha} (\epsilon_{3^\dag,\alpha}^4)_n
  \, (X_{3^\dag,\alpha}^4)^*_n(\mu) \, ( X_{3^\dag,\alpha}^4)_n^*(\mub) \,.
\end{align}  
where we have used $\aver{(\mu \cdot\mub \cdot\Phi_\alpha)^\dag \Phi_3^\nodag \Phi_4^\nodag} = \aver{\Phi_4^\dag \Phi_3^\dag (\mu \cdot\mub \cdot\Phi_\alpha)}^*$. We get the decomposition
\begin{equation} \label{eq:block-decomposition}
  G(z,\zb) =\sum_{\alpha} \sum_{m=1}^{\Nc_{2\alpha}^1} \sum_{n=1}^{\Nc_{34}^{\alpha}}
  (\epsilon_{2\alpha}^1)_m \, (\epsilon_{3^\dag,\alpha}^4)_n \,
  \mathcal{F}_{\alpha,m,n}(z) \, \bar{\cal F}_{\alpha,m,n}(\zb) \,,
\end{equation}
where the conformal blocks are defined by
\begin{align}
  \mathcal{F}_{\alpha,m,n}(z) &= \sum_{\mu} z^{-h_{34}^\alpha+|\mu|}
  \, (X_{2\alpha}^1)_m(\mu) \, (X_{3^\dag,\alpha}^4)_n^*(\mu) \,, \\
  \bar{\cal F}_{\alpha,m,n}(\zb) &= \sum_{\mub} \zb^{-h_{34}^\alpha+|\mub|}
  \, (X_{2\alpha}^1)_m(\mub) \, (X_{3^\dag,\alpha}^4)_n^*(\mub) \,,
\end{align}
and the sums run over orthonormal bases $\{\mu\ket{\Phi_\alpha}\}$ and $\{\mub\ket{\Phi_\alpha}\}$ of $V_\alpha$ and $\Vb_{\alpha^*}$, respectively.
\bigskip

Note that each of the conformal blocks $\mathcal{F}_{\alpha,m,n}(z)$ has the form of an integer power series in $z$, multiplied by a factor $z^{-h_{34}^\alpha}$. The conformal block decomposition \eqref{eq:block-decomposition} is indexed by the invariant primary operators $\Phi_\alpha$ under the $A_N \oplus \Ab_N$ neutral algebra. Each internal primary state $\ket{\Phi_\alpha}$ contributes with a multiplicity (\textit{i.e.} number of independent conformal blocks) given by the product of fusion numbers $\Nc_{2\alpha}^1 \times \Nc_{34}^\alpha$.

\section{Modular properties and Verlinde's formula}
\label{sec:modular}

In this section, we recall the modular properties of the $\mathbb{Z}_N$ orbifold and obtain the orbifold fusion rules from the Verlinde formula under the assumption that $N$ is prime.

\subsection{Torus partition function}
\label{sec:torus_partition_function}

Consider the torus of modular parameter $\tau$
\begin{equation}
  \Tbb_\tau = \Cbb / (\Zbb + \tau\Zbb) \,,
\end{equation}
with $\Im\, \tau >0$, and let us use the notations $q=e^{2i\pi\tau}$ and $\qb=e^{-2i\pi\taub}$.
We denote by $Z_{m,n}(\tau)$ the partition function of $\M_N$, where the copy $a$ is connected to the copy $a+m$ (resp. $a+n$) along the cycle $z \to z+\tau$ (resp. $z \to z+1$). Since changing the orientation of both cycles yields the same torus, clearly $ Z_{m,n}(\tau)= Z_{-m,-n}(\tau)$. Furthermore, the elementary modular transformations act as:

\begin{equation}
  Z_{mn}(\tau+1) = Z_{m-n,n}(\tau) \,,
  \qquad Z_{mn}(-1/\tau) = Z_{n,-m}(\tau) \,.
\end{equation}

For a general modular transformation, we have:

\begin{equation} \label{eq:modular-Zmn}
  Z_{mn} \left( \tau \right) = Z_{am+bn,cm+dn}\left( \frac{a\tau+b}{c\tau+d} \right) \,,
  \qquad (a,b,c,d) \in \Zbb^4 \,, \qquad ad-bc=1 \,.
\end{equation}

The full partition function of the orbifold CFT is defined as the sum over all possible defects:
\begin{equation}
\label{eq_partition_function}
  Z_{\rm orb}(\tau) = \frac{1}{N} \sum_{m,n=0}^{N-1} Z_{mn}(\tau) \,,
\end{equation}
and it is manifestly modular invariant.

Denoting by $Z(\tau)$ the mother CFT partition function, we have the identities
\begin{equation}
  Z_{m0}(\tau) = Z(N\tau) \,,
  \qquad Z_{0m}(\tau) = Z(\tau/N) \,,
  \qquad m \neq 0 \mod N \,.
\end{equation}
Using these, together with the relations \eqref{eq:modular-Zmn}, we can express $Z_{\rm orb}(\tau)$ as:
\begin{align}
  Z_{\rm orb}(\tau) &= \frac{1}{N} \left[
    Z_{00}(\tau) + \sum_{m=1}^{N-1} Z_{m0}(\tau) + \sum_{m=0}^{N-1} \sum_{n=1}^{N-1} Z_{mn}(\tau) \right] \nn \\
  &= \frac{1}{N} Z(\tau)^N + \frac{N-1}{N} \left[
    Z(N\tau) + \sum_{k=0}^{N-1} Z \left(\frac{\tau+k}{N} \right) \right] \,,
\end{align}
which is the standard result \cite{klemm_orbifolds_1990} for the partition function of the $\mathbb{Z}_N$ orbifold.
Recall the definition \eqref{eq:Pi} for the cyclic permutation of copies, and the family of projectors
\begin{equation}
  \Pc_r = \frac{1}{N} \sum_{a=0}^{N-1} \omega^{-ar} \, \Pi^a \,,
\end{equation}
on the eigenspaces of eigenvalue $\omega^r$ of $\Pi$.
By construction, each individual twisted partition function $Z_{mn}(\tau)$ reads
\begin{equation}
  Z_{mn}(\tau) = \Tr_{\mathcal{H}^{[n]}} \left[ \Pi^m \, q^{L_0^{(0)}-Nc/24} \, \qb^{\Lb_0^{(0)}-Nc/24} \right] \,,
\end{equation}
where $\mathcal{H}^{[n]}$ is the space of $\M_N$ states with twist charge $n$. Therefore, the orbifold partition function \eqref{eq_partition_function} can be rewritten as
\begin{equation}
  Z_{\rm orb}(\tau) = \sum_{n=0}^{N-1} \Tr_{\mathcal{H}^{[n]}} \left[ \Pc_0 \, q^{L_0^{(0)}-Nc/24} \, \qb^{\Lb_0^{(0)}-Nc/24} \right] \,.
\end{equation}
Hence we get the simple identity
\begin{equation} \label{eq:Zorb}
  Z_{\rm orb}(\tau) = \Tr_{\mathcal{H}_0} \left[ q^{L_0^{(0)}-Nc/24} \, \qb^{\Lb_0^{(0)}-Nc/24} \right] \,,
\end{equation}
where $\mathcal{H}_0$ is the invariant Hilbert space defined in Sec.~\ref{sec:Hilbert}. We recover that local fields coincide with invariant fields. 

\subsection{Modular characters}

Let us first fix the notations for the characters of the mother CFT, in the case when it is rational, \textit{i.e.} with a finite set of primary operators $\{\phi_1, \dots, \phi_M\}$.
We denote by $\chi_j$ the Virasoro character of the module associated to $\phi_j$ in the mother CFT:
\begin{equation}
  \chi_j(\tau) = \Tr_{v_j} \left( q^{L_0-c/24} \right) \,.
\end{equation}
The $\chi_j$'s transform under the elementary modular maps as
\begin{equation}
  \chi_j(\tau+1) = t_j \, \chi_j(\tau) \,,
  \qquad \chi_i(-1/\tau) = \sum_{j=1}^M S_{ij} \, \chi_j(\tau) \,,
\end{equation}
where $t_j= \exp[2i\pi(h_j-c/24)]$, and $S$ is a unitary matrix. Moreover, since we assume that the mother theory $\M$ is diagonal, the matrix $S$ is real symmetric, and satisfies $(ST)^3=\id$, where $T=\mathrm{diag}(t_1, \dots, t_M)$.
\bigskip

In the orbifold CFT, the characters associated to the $A_N$-modules $V_{J}, V_j^{(r)}, V_j^{[k](r)}$ are, respectively:
\begin{align}
  & \chi_{J}(\tau) = \chi_{j_1}(\tau) \dots \chi_{j_N}(\tau) \,, \\
  & \chi_j^{(r)}(\tau) = \frac{1}{N} \left[ \chi_j(\tau)^N + (N\delta_{r0}-1) \chi_j(N\tau) \right] \,, \\
  & \chi_j^{[k](r)}(\tau) = \frac{1}{N} \sum_{n=0}^{N-1} t_j^{-n/N} \, \omega^{-krn} \, \chi_j\left(\frac{\tau+n}{N} \right) \,.
\end{align}
The proof for these expressions is given in \cite{borisov_systematic_1998}.
and they can be considered as a special case of \cite{Bantay:1997ek,bantay_permutation_2002} (which are also employed in \cite{Kac_2004} and \cite{dong_s-matrix_2021})

From the decomposition of the invariant Hilbert space $\mathcal{H}_0$, we can write the partition function \eqref{eq:Zorb} as
\begin{equation}
  Z_{\rm orb}(\tau) = \sum_{J = [j_1 \dots j_N]} |\chi_{J}|^2
  + \sum_j \sum_{r=0}^{N-1} |\chi_j^{(r)}|^2
  + \sum_j \sum_{k=1}^{N-1} \sum_{r=0}^{N-1} |\chi_j^{[k](r)}|^2 \,,
\end{equation}
where the notations for the sums are the same as in \eqref{eq:H0}.
The orbifold characters transform under the elementary modular maps as
\begin{equation} \label{eq:modular-maps-chi-alpha}
  \chi_\alpha(\tau+1) = \Tc_\alpha \, \chi_\alpha(\tau) \,,
  \qquad \chi_\alpha(-1/\tau) = \sum_\beta \Sc_{\alpha\beta} \, \chi_\beta(\tau) \,,
\end{equation}
with
\begin{equation} \label{eq:T-matrix}
  \Tc_{J} = t_{j_1}\dots t_{j_N} \,,
  \qquad \Tc_j^{(r)} = t_j^N \,,
  \qquad \Tc_j^{[k](r)} = \omega^{kr} \, t_j^{1/N} \,,
\end{equation}
and
\begin{equation} \label{eq:S-matrix}
  \begin{aligned}
    & \Sc_{I,J} = \sum_{a=0}^{N-1} S_{i_1,j_{1+a}} \dots S_{i_N,j_{N+a}} \,, \\
    & \Sc_{I, j^{(r)}} = \Sc_{j^{(r)},[i_1 \dots i_N]} = S_{i_1,j} \dots S_{i_N,j} \,, \\
    & \Sc_{i^{(r)},j^{(s)}} = \frac{S_{ij}^N}{N} \,, \\
    & \Sc_{i^{(r)},j^{[k](s)}} = \Sc_{j^{[k](s)},i^{(r)}} = \frac{\omega^{-kr} \, S_{ij}}{N} \,, \\
    & \Sc_{i^{[k](r)}, j^{[\ell](s)}} = \frac{\omega^{-ks-\ell r} \, (P_{\ell.k^{-1}})_{ij}}{N} \,, \\
    & \Sc_{I,j^{[k](r)}} = \Sc_{j^{[k](r)},I} = 0 \,.
  \end{aligned}
\end{equation}
The matrices $P_n$ appearing in this expression are defined as
\begin{equation} \label{eq:Pn}
  P_n = T^{-n/N} \cdot Q_n \cdot T^{[[-n^{-1}]]/N} \,, \qquad n \in \Zbb_N^\times \,,
\end{equation}
where $[[-n^{-1}]]$ denotes the inverse of $(-n)$ modulo $N$, with $0<[[-n^{-1}]]<N$, and $Q_n$ is the matrix representing the linear action of the modular map
\begin{equation} \label{eq:qn}
  \tau \mapsto q_n(\tau) = \frac{n\tau -(n [[-n^{-1}]]+1)/N}{N\tau - [[-n^{-1}]]} \,
\end{equation}
on the characters $\chi_j$ of the mother CFT.
We introduce the conjugation matrix $\Cc$, with matrix elements
\begin{equation}
  \Cc_{\alpha \beta} = \delta_{\alpha,\beta^\dag} \,,
\end{equation}
and the diagonal matrix $\Tc$ with matrix elements $\Tc_\alpha$.
The orbifold modular matrices $\Sc$ and $\Tc$ satisfy the properties
\begin{align}
  & \Sc^t = \Sc \,, \label{eq:St=S} \\
  & \Sc^2 = \Cc \,, \\
  & \Sc^\dag \Sc = \id \,, \\
  & (\Sc \Tc)^3 = \Cc \label{eq:(ST)^3=C} \,,
\end{align}
where $\Sc^t$ denotes the transpose of $\Sc$.

The main arguments for the proofs of these properties are given in the Appendix~\ref{sec:proof-modular}.
Note that the relations \eqref{eq:modular-maps-chi-alpha} are not sufficient to determine completely the matrices $\Sc$ and $\Tc$, because some distinct $A_N$-modules have the same characters, namely
\begin{align}
  & \chi_{[j_1 \dots j_N]}(\tau) = \chi_{[j_{p(1)} \dots j_{p(N)}]}(\tau)
  \quad \text{for any permutation $p$ of $\{1,\dots,N\}$,} \\
  & \chi_j^{(r)}(\tau) = \chi_j^{(s)}(\tau) \quad \text{if } r,s \neq 0 \,, \\
  & \chi_j^{[k](r)}(\tau) = \chi_j^{[\ell](s)}(\tau) \quad \text{if } rk = s\ell \mod N \,.
\end{align}
However, the expressions (\ref{eq:T-matrix}--\ref{eq:S-matrix}) define a simple and elegant solution to the constraints (\ref{eq:St=S}--\ref{eq:(ST)^3=C}) in terms of the modular matrices of the mother CFT. Furthermore, it agrees with the $S$ matrix given in equation (10) of \cite{bantay_permutation_2002}. 

\subsection{Fusion rules}
\label{sec:fusion-rules}

Recall our notation for the fusion numbers:
\begin{equation}
  \Phi_\alpha \times \Phi_\beta \to \sum_\gamma \Nc_{\alpha\beta}^\gamma  \, \Phi_\gamma \,,
\end{equation}
where $\Phi_\alpha,\Phi_\beta,\Phi_\gamma$ are invariant primary operators.
The fusion numbers can be computed through the Verlinde formula
\begin{equation}
  \Nc_{\alpha\beta}^\gamma = \sum_\delta \frac{\Sc_{\alpha\delta}^\nodag \, \Sc_{\beta\delta}^\nodag \, \Sc_{\gamma\delta}^\dag}{\Sc_{\id\delta}} \,,
\end{equation}
where the sum runs over all possible invariant primary operators $\Phi_\delta$ described in Section~\ref{sec:symm-operators}. This formula, together with (\ref{eq:St=S}--\ref{eq:(ST)^3=C}), directly ensures that the properties are satisfied
\begin{equation}
  \Nc_{\alpha\beta}^\gamma = \Nc_{\beta\alpha}^\gamma \,,
  \qquad \Nc_{\alpha\beta}^\gamma = \Nc_{\alpha\gamma^\dag}^{\beta^\dag} \,,
\end{equation}
are satisfied.
In the case of untwisted operators, we get
\begin{align}
  & \Nc_{I,J}^K= \sum_{a,b=0}^{N-1} 
  N_{i_{1+a},j_{1+b}}^{k_1} \dots N_{i_{N+a},j_{N+b}}^{k_N} \,, \label{eq:fusionNDNDND} \\
  & \Nc_{I,J}^{k^{(r)}} = \sum_{a=0}^{N-1} N_{i_{1+a},j_1}^k \dots N_{i_{N+a},j_N}^k \,, \label{eq:fusionNDNDD} \\
  & \Nc_{I,j^{(r)}}^{k^{(s)}} = N_{i_1,j}^k \dots N_{i_N,j}^k \,, \label{eq:fusionNDDD} \\
  & \Nc_{i^{(r)},j^{(s)}}^{k^{(t)}} = \delta_{r+s,t} \, N_{ij}^k \,, \label{eq:fusionDDD}
\end{align}
which confirms the hypothesis that the OPE coefficients constructed in Sec.~\ref{sec:correlation-functions} span the solution space of the linear system of equations which derive from the orbifold algebraic rules and Ward identities.
The fusion numbers involving twist operators are given by
\begin{align}
  & \Nc_{i^{[p](r)} j^{[q](s)}}^{K} = \delta_{p+q,0} \, \sum_{\ell=1}^M \frac{S_{i\ell}S_{j\ell} \cdot S_{k_1\ell}\dots S_{k_N\ell}}{S_{1\ell}^N} \,, \\
  & \Nc_{i^{[p](r)} j^{[q](s)}}^{k^{(t)}} =  \frac{\delta_{p+q,0}}{N} 
  \sum_{\ell=1}^M \left[ \frac{S_{i\ell}S_{j\ell}S_{k\ell}^N}{S_{1\ell}^N}
  + \sum_{n=1}^{N-1} \omega^{np(r+s-t)} \frac{(P_{-n})_{i\ell}(P_{n})_{j\ell} S_{k\ell}}{S_{1\ell}} \right] \label{eq:Nttd} \,, \\
  & \Nc_{i^{[p](r)} j^{[q](s)}}^{k^{[m](t)}} =  \frac{\delta_{p+q,m}}{N}
    \sum_{\ell=1}^M \left[
    \frac{S_{i\ell}S_{j\ell}S_{k\ell}}{S_{1\ell}^N} + \sum_{n=1}^{N-1} \omega^{n(r+s-t)}
    \frac{(P_{pn^{-1}}^\dag)_{i\ell} (P_{qn^{-1}}^\dag)_{j\ell} (P_{mn^{-1}}^\nodag)_{k\ell}}{S_{1\ell}}
    \right] \,. \label{eq:Nttt}
\end{align}
The fusion numbers~\eqref{eq:Nttd} depend on the Fourier indices $r,s,t$ only through the combination $(r+s-t)$, as expected from the discussion in Sec.~\ref{sec:correlation-functions}.
Using the unitarity of $S$, some of the terms can be expressed using the fusion numbers of the mother CFT:
\begin{align}
  & \sum_{\ell=1}^M \frac{S_{i\ell}S_{j\ell} \cdot S_{k_1\ell}\dots S_{k_N\ell}}{S_{1\ell}^N}
    = \sum_{\ell_1,\dots,\ell_{N-1}=1}^M N_{ij}^{\ell_1}
    \times N_{\ell_1 k_1}^{\ell_2} N_{\ell_2 k_2}^{\ell_3} \dots N_{\ell_{N-2} k_{N-2}}^{\ell_{N-1}}
    \times N_{\ell_{N-1} k_{N-1}}^{k_N} \,, \\
  & \sum_{\ell=1}^M \frac{S_{i\ell}S_{j\ell}S_{k\ell}^N}{S_{1\ell}^N}
    = \sum_{\ell_1,\dots,\ell_{N-1}=1}^M N_{ij}^{\ell_1}
    \times N_{\ell_1 k}^{\ell_2} N_{\ell_2 k}^{\ell_3} \dots N_{\ell_{N-2} k}^{\ell_{N-1}}
    \times N_{\ell_{N-1} k}^k \,.
\end{align}
This is also the case for the first term in \eqref{eq:Nttt}. For instance, for $N=3$, we have
\begin{equation}
  \sum_{\ell=1}^M \frac{S_{i\ell}S_{j\ell}S_{k\ell}}{S_{1\ell}^3}
  = \sum_{\ell_1,\ell_2,\ell_3=1}^M N_{i\ell_1}^{\ell_2} N_{j\ell_2}^{\ell_3} N_{k\ell_3}^{\ell_1} \,,
\end{equation}
and similar but more complicated expressions hold for $N > 3$.

\section{Example applications}
\label{sec:examples}

In this section, we shall present some applications of our results on  the $\Zbb_3$ orbifolds of minimal CFTs.

\subsection{The $\Zbb_3$ orbifold of the Yang-Lee CFT}
In this section, we shall present a simple application of our results for the operator algebra of the cyclic orbifold CFT. We choose as mother theory the Yang-Lee minimal model $\mathcal{M}(5,2)$ with central charge $c=-22/5$, and primary operator content $\{\id,\phi\}$ with conformal dimensions $h_{\id}=0$ and $h_{\phi}=-1/5$. The modular $S$-matrix associated to this CFT is given by:
\begin{equation}
   S=\frac{2}{\sqrt{5}}\left(\begin{array}{cc}
    -\sin(2\pi/5)  & \sin(4\pi/5) \\
    \sin(4\pi/5)   &  \sin(2\pi/5)
   \end{array}\right) \,.
\end{equation}
The non-trivial fusion rule in the Yang-Lee CFT is
\begin{equation}
  \phi \times \phi \to \id + \phi \,.
\end{equation}
We will consider the $\mathbb{Z}_3$ cyclic orbifold of the above CFT. Its primary operator content (with respect to the neutral algebra $A_N\oplus \Ab_N$) consists of:
\begin{itemize}
\item 2 non-diagonal untwisted operators: $[\id,\id,\phi]$ and $[\id,\phi,\phi]$,
\item 6 diagonal untwisted operators: $\id^{(r)}, \Phi^{(r)}$ with $r \in \Zbb_3$,
\item 12 twist operators: $\sigma_\id^{(r)}, \sigma_\id^{\dag(r)}, \sigma_\phi^{(r)} \sigma_\phi^{\dag(r)}$ with $r \in \Zbb_3$.
\end{itemize}
Recall the definition of non-diagonal untwisted operators, e.g.
\begin{equation}
  [\id,\id,\phi] := \frac{1}{\sqrt 3} (\id\otimes\id\otimes\phi+\id\otimes\phi\otimes\id+\phi\otimes\id\otimes\id) \,.
\end{equation}

Using the results of Sec.~\ref{sec:fusion-rules}, we get the fusion rules between untwisted operators:
\begin{equation} \label{eq:fusionYL1}
  \begin{aligned}
    & \repND{\id}{\id}{\phi}\times \repND{\id}{\id}{\phi}
    \rightarrow \repND{\id}{\id}{\phi}+ 2\repND{\id}{\phi}{\phi}+ \sum_{r \in \Zbb_3} \repD{\id}{r} \,, \\
    & \repND{\id}{\id}{\phi}\times \repND{\id}{\phi}{\phi}
    \rightarrow \repND{\id}{\id}{\phi}+ 2\repND{\id}{\phi}{\phi}+ \sum_{r \in \Zbb_3} \repD{\Phi}{r} \,, \\
    & \repND{\id}{\phi}{\phi}\times \repND{\id}{\phi}{\phi}
    \rightarrow 2 \repND{\id}{\id}{\phi}+ 3\repND{\id}{\phi}{\phi}+ \sum_{r \in \Zbb_3} \repD{\id}{r}+2\sum_{r \in \Zbb_3} \repD{\Phi}{r} \,, \\
    & \repND{\id}{\id}{\phi}\times \repD{\id}{r}
    \rightarrow  \repND{\id}{\id}{\phi} \,, \\
    & \repND{\id}{\phi}{\phi}\times \repD{\id}{r} \rightarrow  \repND{\id}{\phi}{\phi} \,, \\
    & \repD{\id}{r}\times \repD{\id}{s} \rightarrow \repD{\id}{r+s} \,, \\
    & \repD{\Phi}{r}\times \repD{\id}{s} \rightarrow \repD{\Phi}{r+s} \,, \\
    & \repND{\id}{\id}{\phi}\times \repD{\Phi}{r}
    \rightarrow \repND{\id}{\phi}{\phi}+\sum_{s \in \Zbb_3} \repD{\Phi}{s} \,, \\
    & \repND{\id}{\phi}{\phi}\times \repD{\Phi}{r}
    \rightarrow  \repND{\id}{\id}{\phi}+ 2 \repND{\id}{\phi}{\phi}+\sum_{s \in \Zbb_3} \repD{\Phi}{s} \,, \\
    & \repD{\phi}{r}\times \repD{\phi}{s}
    \rightarrow  \repND{\id}{\id}{\phi}+ \repND{\id}{\phi}{\phi} +2\times\repD{\id}{r+s}+\repD{\Phi}{r+s} \,.
  \end{aligned}
\end{equation}
The fusion rules between untwisted and twisted operators read
\begin{equation} \label{eq:fusionYL2}
  \begin{aligned}
    & \repND{\id}{\id}{\phi} \times \repT{\id}{s}
    \rightarrow \sum_r \repT{\phi}{r} \,, \\
    & \repND{\id}{\phi}{\phi} \times \repT{\id}{s}
    \rightarrow \sum_r \repT{\id}{r}+\sum_{r \in \Zbb_3} \repT{\phi}{r} \,, \\
    & \repND{\id}{\id}{\phi} \times \repT{\phi}{s}
    \rightarrow \sum_r \repT{\id}{r}+\sum_{r \in \Zbb_3} \repT{\phi}{r} \,, \\
    & \repND{\id}{\phi}{\phi} \times \repT{\phi}{s}
    \rightarrow \sum_r \repT{\id}{r} + 2\sum_{r \in \Zbb_3} \repT{\phi}{r} \,,\\
    & \repD{\id}{r}\times \repT{\id}{s} \rightarrow \repT{\id}{s+r} \,, \\
    & \repD{\Phi}{r}\times \repT{\id}{s}
    \rightarrow \repT{\id}{s+r}+\repT{\phi}{s+r}+\repT{\phi}{s+r+1} \,, \\
    & \repD{\id}{r}\times \repT{\phi}{s} \rightarrow \repT{\phi}{s+r} \,, \\
    & \repD{\Phi}{r}\times \repT{\phi}{s} \rightarrow
    \repT{\id}{s+r-1}+\repT{\id}{s+r}+\sum_{t \in \Zbb_3} \repT{\phi}{t} \,.
  \end{aligned}
\end{equation}
Finally, the fusion rules between twist operators read
\begin{equation} \label{eq:fusionYL3}
  \begin{aligned}
    & \repT{\id}{r} \times \repTdag{\id}{s}
    \rightarrow \repND{\id}{\phi}{\phi} +\repD{\id}{s+r}+\repD{\Phi}{s+r} \,, \\
    & \repT{\phi}{r} \times  \repTdag{\id}{s}
    \rightarrow \repND{\id}{\id}{\phi}+ \repND{\id}{\phi}{\phi} +\repD{\Phi}{s+r-1}+\repD{\Phi}{s+r} \,,  \\
    & \repT{\phi}{r} \times \repTdag{\phi}{s}
    \rightarrow \repND{\id}{\id}{\phi}+ 2\repND{\id}{\phi}{\phi} +\repD{\id}{s+r}+\sum_{r \in \Zbb_3} \repD{\Phi}{r} \,, \\
    & \repT{\id}{r} \times \repT{\id}{s}
    \rightarrow \repTdag{\id}{r+s}+\repTdag{\id}{r+s+1} +\repTdag{\phi}{r+s} \,, \\
    & \repT{\phi}{r} \times \repT{\id}{s}
    \rightarrow \repTdag{\id}{r+s} + \sum_{t \in \Zbb_3}\repTdag{\phi}{t} \,, \\
    & \repT{\phi}{r} \times \repT{\phi}{s}
    \rightarrow \sum_{t \in \Zbb_3} \repTdag{\id}{t} + \repTdag{\phi}{r+s} + \repTdag{\phi}{r+s+1}
    + 2 \repTdag{\phi}{r+s+2} \,.
  \end{aligned}
\end{equation}

The other fusion rules are easily obtained by symmetry under conjugation.

As pointed out in Sec.~\ref{sec:correlation-functions}, one may treat the OPE of twist operators $\sigma_i^{(r)} \times \sigma_j^{\dag(s)}$ or $\sigma_i^{(r)} \times \sigma_j^{(s)}$, by ``unfolding'' the three-point correlator corresponding to a particular orbifold fusion rule, as a mother CFT correlator. This is done in two steps: first, one translates the orbifold correlation function into an expectation value on the replicated surface $\Sigma_n$. Then, one conformally maps this quantity to a correlator defined on a Riemann surface that is more amenable to calculations (the Riemann sphere $\Cbb$ in our first example).
This approach is exemplified in Appendix~\ref{sec:proof-3pt} for the calculation of the orbifold structure constant 
\begin{equation}
  \mathcal{C}^{[\id,\phi,\phi]}_{\sigma_{\phi}^{(0)},\sigma_{\phi}^{\dag(0)}}
  = \aver{[\id,\phi,\phi] \cdot \sigma_{\phi}^{(0)} \cdot \sigma_{\phi}^{\dag(0)}} \neq 0 \,.
\end{equation}
Since the structure constant is non-vanishing, we can conclude that indeed the OPE $\sigma_{\phi}^{(0)} \times \sigma_{\phi}^{\dag(0)}$ produces the module $[\id,\phi,\phi]$.
However, this unfolding does not allow us to extract the value of the associated multiplicity. From \eqref{eq:fusionYL3}, we see that this multiplicity is two. Furthermore, using this technique to infer the fusion rules of the cyclic orbifold CFT from the mother CFT data is only feasible provided the resulting correlator can be calculated, and the unfolding map is known. 

One encounters both of these difficulties in trying to find the fusion rules between twist operators in the same twist charge sector.
Let us consider the following orbifold three-point function of twist operators:
\begin{equation}
  \mathcal{C}_{\sigma_j \sigma_k}^{\sigma_i^\dag}
  = \aver{\sigma_i \cdot \sigma_j \cdot \sigma_k} \,.
\end{equation}
This translates into a three-point function on the three-sheeted cover of the Riemann sphere, with  branch points at $(0,1,\infty)$, which we denote by $\Sigma_3$:
\begin{equation} \label{eq:correlatorreplicated}
    \aver{\phi_i \cdot \phi_j \cdot \phi_k}_{\Sigma_3} \,.
\end{equation}
We can now calculate the genus of this surface $\Sigma_3$, using the Riemann-Hurwitz formula
\begin{equation} \label{eq:RiemannHurwitz}
  2g-2 = N(2h-2) + \sum_{i=1}^p (e_i-1) \,,
\end{equation}
giving the genus $g$ of an $N$-sheeted cover of a surface of genus $h$, with $p$ branch points having ramification indices $e_i$. In our case, all the branch points have ramification indices $e_i=3$, so that we find $g=1$. Thus, there exists a conformal map $z\mapsto t(z)$ between the surface $\Sigma_3$ and a torus $\mathbb{T}_\tau$  which allow us to relate \eqref{eq:correlatorreplicated} to the following three point function on the torus
\begin{equation}\label{eq:correlatoronthetorus}
  \langle \phi_i(t_1,\bar{t}_1)\phi_j(t_2,\bar{t}_2)\phi_k(t_3,\bar{t}_3)\rangle_{\mathbb{T}_\tau} \,.
\end{equation}
To complete such a calculation, one would have two non-trivial problems to solve: finding the conformal map $z\mapsto t(z)$, and calculating the three-point correlator \eqref{eq:correlatoronthetorus}. Assuming the conformal map has been found, one needs to calculate the torus correlators on a case-by-case basis. For example, when $(\phi_i,\phi_j,\phi_k)=(\id,\id,\id)$ or $(\id,\id,\phi)$, results are already known -- they correspond respectively to the partition function of the Yang-Lee CFT, and the torus one-point function $\aver{\phi}_{\mathbb{T}_\tau}$, which have been calculated in \cite{gaberdiel_modular_2008}.

However, the issue of counting multiplicities still remains, and as we go beyond this relatively simple example of the $\Zbb_3$ orbifold of the Yang-Lee CFT, to higher $\Zbb_N$ orbifolds, the formula \eqref{eq:RiemannHurwitz} shows that unfolding would require the calculation of correlators on surfaces of genus $g\geq 2$, for which few exact results are available in the literature.

\subsection{Counting conformal blocks of twist correlators}
We will provide a few examples of conformal block counting  for four-point correlators in $\Zbb_3$ cyclic orbifolds.
\bigskip

\paragraph{The $\Zbb_3$ cyclic orbifold of the Ising CFT.} We will first illustrate this by using the minimal model $\mathcal{M}(4,3)$ (Ising CFT) as mother theory, with central charge $c=1/2$, and primary operator content $\{\id,s,\varepsilon\}$ with conformal dimensions $h_{\id}=0$, $h_s=1/16$, $h_{\varepsilon}=1/2$. To avoid confusion with the twist operators, we d
The modular $S$-matrix associated to this CFT is given by:
\begin{equation}
  S=\frac{1}{2}\left(\begin{array}{ccc}
    1 & 1 & \sqrt{2} \\
    1 & 1 & -\sqrt{2} \\
    \sqrt{2} & -\sqrt{2} & 0
  \end{array}\right) \,,
\end{equation}
and the fusion rules are
\begin{equation}
  s \times s \to \id + \varepsilon \,,
  \qquad s \times \varepsilon \to s \,,
  \qquad \varepsilon\times \varepsilon \to \id \,.
\end{equation}
We remind that, for a  four-point function $\aver{\Phi_1\Phi_2\Phi_3\Phi_4}$, the dimension of its space of conformal blocks can be calculated from the fusion numbers of the theory \cite{di_francesco_conformal_1997}:
\begin{equation}
  \mathcal{D}_{\Phi_1,\Phi_2}^{\Phi_3,\Phi_4} =
  \sum_\alpha \mathcal{N}_{\Phi_1,\Phi_2}^{\Phi_\alpha} \mathcal{N}_{\Phi_\alpha,\Phi_3}^{\Phi_4} \,,
\end{equation}
where the sum on $\alpha$ runs over all the primary fields in the CFT.
Using the above, we will calculate the dimensions of the space of conformal blocks for a few correlators of physical interest in the $\Zbb_3$ orbifold of the Ising CFT.

Let us first  consider the four-point functions of twist operators:
\begin{equation}
    \aver{\sigma_j(z_1,\zb_1) \sigma^\dag_k(z_2,\zb_2)\sigma_j(z_3,\zb_3)\sigma^\dag_k(z_4,\zb_4)} \,,
\end{equation}
with $j,k \in\{\id,\varepsilon\}$. In this case, one finds:
\begin{equation}
  \boxed{
    \mathcal{D}_{\sigma_j,\sigma^\dag_k}^{\sigma_j,\sigma^\dag_k}
    = \sum_\alpha \left(\Nc_{\sigma_j,\sigma^\dag_k}^{\Phi_\alpha}\right)^2 = 4 \,,
  }
\end{equation}
where the sum runs over the untwisted operators.
The correlator with $j=k=\id$ gives the leading universal contribution in the calculation of the third R\'enyi entropy of two disjoint intervals in the ground state of a quantum spin chain with periodic boundary conditions, while the correlators with composite twist insertions (i.e. for which $j,k$ can be $\varepsilon$) provide expressions for the universal parts of the first two subleading terms, as per the reasoning of \cite{dupic_entanglement_2018}.

Also of physical relevance are correlators of the type:
\begin{equation}
  \aver{\sigma_j(z_1,\zb_1) \sigma^\dag_j(z_2,\bar{z}_2)\Phi^{(0)}_i(z_3,\zb_3)\Phi_i^{(0)}(z_4,\zb_4)} \,,
\end{equation}
with $j\in\{\id,\varepsilon\}$ and $i \in \{s,\epsilon\}$. These allow for the calculation of leading and subleading contributions to the third R\'enyi entropy of a single interval in an excited state of the quantum Ising chain with periodic boundary conditions.
We find:
\begin{equation}
  \mathcal{D}_{\sigma^\nodag_\id,\sigma^\dag_\id}^{\varepsilon^{(0)},\varepsilon^{(0)}}
  = \mathcal{D}_{\sigma^\nodag_\varepsilon,\sigma^\dag_\varepsilon}^{\varepsilon^{(0)},\varepsilon^{(0)}}
  = \mathcal{D}_{\sigma^\nodag_\id,\sigma^\dag_\varepsilon}^{s^{(0)},s^{(0)}} = 1 \,,
\end{equation}
and
\begin{equation}
 \mathcal{D}_{\sigma^\nodag_\id,\sigma^\dag_\id}^{s^{(0)},s^{(0)}}
 = \mathcal{D}_{\sigma^\nodag_\varepsilon,\sigma^\dag_\varepsilon}^{s^{(0)},s^{(0)}} = 2 \,.
\end{equation}

\paragraph{The $\Zbb_3$ orbifold of a Virasoro minimal model $\mathcal{M}_{p,q}$.}
One can furthermore provide numerical checks\footnote{if the number of primary fields of the mother CFT has a reasonable value} for the claims of \cite{dupic_entanglement_2018} at $N=3$.

Provided the mother CFT is a diagonal Virasoro minimal model $\mathcal{M}_{p,q}$ \cite{di_francesco_conformal_1997}, 
the cyclic orbifold correlation function:
\begin{equation}\label{eq:correlatorblockdimensions}
  \aver{\sigma_\id(z_1,\zb_1),\sigma^\dag_\id(z_2,\zb_2),\Phi^{(0)}_{21}(z_3,\zb_3),\Phi_{21}^{(0)}(z_4,\zb_4)} \,,
\end{equation}
where $\phi_{21}$ is the primary field with Kac indices $(m,n)=(2,1)$ (see \cite{di_francesco_conformal_1997}). According to \cite{dupic_entanglement_2018}, this correlator should satisfy a third order differential equation. 
We've found evidence for this claim by calculating the dimension of the space of conformal blocks of this correlator for a few minimal models. The results are given in Table \ref{tab:minimalmodeldimensions}.
\begin{table}[h]
    \centering
    \begin{tabular}{|c|c|c|c|c|c|c|}
    \hline
     $\mathcal{M}(p,q)$ & $ \mathcal{M}(4,3)$  & $ \mathcal{M}(5,4) $ & $\mathcal{M}(6,5) $ & $\mathcal{M}(7,5) $ &  $\mathcal{M}(7,6) $ &  $\mathcal{M}(8,7)$ \\
      \hline
    $ \mathcal{D}_{\sigma^\nodag_\id,\sigma^\dag_\id}^{\Phi_{21}^{(0)},\Phi_{21}^{(0)}} $ &  1 & 2 & 2 & 3& 3 & 3\\
    \hline
    \end{tabular}
    \caption{Dimension of the space of conformal blocks for the correlator (\ref{eq:correlatorblockdimensions}) in diverse minimal models $\mathcal{M}(p,q)$. }
    \label{tab:minimalmodeldimensions}
\end{table}
We see that indeed the space of conformal dimensions satisfies the bound
\begin{equation}
  \mathcal{D}_{\sigma^\nodag_\id,\sigma^\dag_\id}^{\Phi_{21}^{(0)},\Phi_{21}^{(0)}} \leq 3 \,,
\end{equation}
conjectured in \cite{dupic_entanglement_2018} for all the cases we've checked. We should mention, that this conjectured third order equation has been derived in \cite{Estienne:2023tdw}.Furthermore, the conformal blocks in the $z_1\rightarrow z_2$ channel were found in  \cite{Estienne:2023tdw} to correspond to the following twist field fusion rules:
\begin{equation}
    \sigma_\id\times \sigma^{\dag}_\id \rightarrow \Phi_{\id}+\Phi_{[\id\,\phi_{1,3}\,\phi_{1,3}]}+\Phi_{1,3}
\end{equation}
in agreement with the results of (\ref{eq:fusionrulesintro}).

In general, knowing the dimension of the space of conformal blocks can be a useful guide in determining the BPZ type differential equation that an orbifold correlator involving twist fields satisfies. This is because conformal blocks form bases of solutions around the singular points of these equations, so their order can be inferred from the dimensions $\mathcal{D}$. This, in turn, provides hints for finding which combination of orbifold null-vectors and Ward identities \cite{dupic_entanglement_2018} one should manipulate in order to recover the BPZ type  equation.

\section{Conclusion}
\label{sec:conclusion}

In this article, we have provided a formal analysis of the $\mathcal{M}_N$ orbifold CFT, with prime $N$, of a diagonal and rational mother CFT $\mathcal{M}$. We have classified the operators of the theory under the neutral algebra of the full orbifold symmetry algebra. Analysing the spectrum of primary operators under $A_N \oplus \Ab_N$, we have found exact solutions for its modular data, as well as closed expressions for the fusion numbers of $\mathcal{M}_N$.
Our results are consistent with the ones obtained for $N=2$ in \cite{borisov_systematic_1998}\footnote{Note that since at $N=2$ there is only one twist charge sector, there are no fusion rules involving three twist fields.}. 

To showcase the use of our results, we have explicitly given all the fusion rules in the $\mathbb{Z}_3$-orbifold of the Yang-Lee CFT. We have also commented on the limitations of the unfolding approach, which requires, even in this relatively simple case, the computation of correlators on surfaces with genus $g\geq 1$. As a second example, we have shown how the fusion numbers can be used to count conformal blocks and commented on applications to the  calculation of R\'enyi entropies. The block counting examples we have presented are consistent with previous results in the literature \cite{dupic_entanglement_2018}. 

Natural directions for investigations can be realized by considering more intricate mother CFTs as starting points. One could consider CFTs which are non-diagonal, non-rational or have extended symmetries as mother CFTs and generalize the results of this paper. Alternatively, one could consider orbifolding by a generic $\mathbb{Z}_N$ group, \emph{i.e.} extending the analysis to non-prime $N$. In this case each twist sector $[k]$ will depend on the greatest common divisor $\mathrm{gcd}(k,N)$. This implies that both the induction procedure and the classification of $A_N\oplus\Ab_N$ primary fields would have to be adapted, as exemplified at the end of Section \ref{sec:symm-operators}.

Finally, it would be mathematically interesting to find an interpretation for the expressions of the fusion numbers \eqref{eq:Nttd} and \eqref{eq:Nttt} in terms of fusion processes in the mother CFT. For $N=2$, this has been done in \cite{pradisi_planar_1995}, and the relevant surface is the crosscap. To our knowledge, generalizing this argument to generic $N$ remains an open problem.

\section*{Appendix}

\appendix

\section{Proofs of the properties of invariant operators}
\label{sec:proof-primary-op}

In this part of the Appendix, we give the proofs for the non-trivial statements made in Sec.~\ref{sec:symm-operators}, about the properties of the operators of the form $\Phi_{[j_1, \dots, j_N]}$, $\Phi_j^{(r)}$ and $\sigma_j^{[k](r)}$ under the neutral algebra $A_N \oplus \Ab_N$. In particular, we say that an operator $\Phi$ is primary under $A_N \oplus \Ab_N$ if and only if
\begin{equation}
  L_{m_1}^{(r_1)} \dots L_{m_p}^{(r_p)} \cdot \Phi = \Lb_{m_1}^{(r_1)} \dots \Lb_{m_p}^{(r_p)} \cdot \Phi = 0
\end{equation}
for any $(m_1, \dots, m_p)$ such that $m_1 \leq \dots \leq m_p$ and $m_1+\dots+m_p>0$, and $r_1+\dots+r_p=0 \mod N$.

\subsection{Untwisted sector}
\label{app:untwistedsector_primary}
The untwisted sector contains local fields in $\M^{\otimes N}$ that are invariant under cyclic permutations, therefore it is spanned by fields of the form 
\begin{equation}
\label{eq_untwisted_generator}
  \Psi = \Pc_0\left[
    L_{m_1}^{(r_1)} \dots L_{m_p}^{(r_p)} \cdot \Lb_{\mb_1}^{(\rb_1)} \dots \Lb_{\mb_\pb}^{(\rb_\pb)}
    \cdot (\phi_{j_1} \otimes \dots \otimes \phi_{j_N}) 
  \right] \,,  \quad \text{with} \ \begin{cases}
  m_1 \leq \dots \leq m_p \leq 0 \,, \\
    \mb_1 \leq \dots \leq \mb_\pb \leq 0 \,.
  \end{cases}
\end{equation}
where $\Pc_0$ is the projector onto the invariant subspace, all $m_i, \mb_i$ are integers, and $\phi_{j_i}$ are primary fields in the mother theory. Let $\Pi$ be the elementary cyclic permutation of the copies, which acts on untwisted operators as
\begin{equation} \label{eq:Pi}
  \Pi \cdot (\phi_{j_1} \otimes \dots \otimes \phi_{j_{N-1}} \otimes \phi_{j_N}) \cdot \Pi^{-1}
  = (\phi_{j_2} \otimes \dots \otimes \phi_{j_N} \otimes \phi_{j_1}) \,.
\end{equation}
and let $  \Pc_s$ be the projectors
\begin{equation}
  \Pc_s[\Phi] = \frac{1}{N} \sum_{a=0}^{N-1} \omega^{-sa} \, \Pi^a \cdot \Phi \cdot \Pi^{-a} \,.
\end{equation}
Since $  \Pi \cdot T^{(r)}(z) \cdot \Pi^{-1} = \omega^r \, T^{(r)}(z)$, where $\omega=\exp(2i\pi/N)$,  it follows that  
\begin{equation}
  \Pi \cdot L_{m}^{(r)} \cdot \Pi^{-1} = \omega^r \,  L_{m}^{(r)} \,.
\end{equation}
and therefore 
\begin{equation}
 L_{m}^{(r)} \Pc_s = \Pc_{r+s} L_{m}^{(r)}
\end{equation}
So the field $\Psi$ from \eqref{eq_untwisted_generator} can be recast as
\begin{equation}
\label{eq_untwisted_generator_2}
  \Psi = L_{m_1}^{(r_1)} \dots L_{m_p}^{(r_p)} \cdot \Lb_{\mb_1}^{(\rb_1)} \dots \Lb_{\mb_\pb}^{(\rb_\pb)}
  \cdot \Pc_{-r-\rb}[\phi_{j_1} \otimes \dots \otimes \phi_{j_N}] \,,
\end{equation}
where $r=r_1+\dots+r_p$ and $\rb=r_1+\dots+\rb_\pb$. We now have to distinguish two cases, depending on whether $\phi_{j_1} \otimes \dots \otimes \phi_{j_N}$ is invariant or not.

\paragraph{Non-diagonal untwisted operators.} When $\phi_{j_1} \otimes \dots \otimes \phi_{j_N}$ is not invariant under cyclic permutations, states of the form \eqref{eq_untwisted_generator} are said to be \emph{non-diagonal}. In particular, the most relevant field in that sector is 
\begin{equation}
\Phi_{[j_1\dots j_N]} \propto \Pc_0 (\phi_{j_1} \otimes \dots \otimes \phi_{j_N}) \,. 
\end{equation}
and therefore it is necessarily primary. We now prove that any operator in the untwisted, non-diagonal sector is an $(A_N \oplus \Ab_N)$-descendant of some primary operator of the form $\Phi_{[j_1 \dots j_N]}$. \\

Let's start from a generic state of the form \eqref{eq_untwisted_generator_2}, that is 
\begin{equation}
  \Psi = L_{m_1}^{(r_1)} \dots L_{m_p}^{(r_p)} \cdot \Lb_{\mb_1}^{(\rb_1)} \dots \Lb_{\mb_\pb}^{(\rb_\pb)}
  \cdot \Pc_{-r-\rb}[\phi_{j_1} \otimes \dots \otimes \phi_{j_N}] \,,
\end{equation}
where $r=r_1+\dots+r_p$ and $\rb=r_1+\dots+\rb_\pb$. Now we simply observe that 
\begin{align}
  L_0^{(r)} \cdot \Pc_s[\phi_{j_1} \otimes \dots \otimes \phi_{j_N}] =   \overline{L}_0^{(r)} \cdot \Pc_s[\phi_{j_1} \otimes \dots \otimes \phi_{j_N}]  = \wh{h}^{(r)}_{j_1\dots j_N} \, \Pc_{s+r}[\phi_{j_1} \otimes \dots \otimes \phi_{j_N}] \,,
\end{align}
where
\begin{equation}
  \wh{h}^{(r)}_{j_1\dots j_N} = \sum_{a=1}^N \omega^{ar} \, h_{j_a} \,.
\end{equation}
The point is that in the non-diagonal sector there exists $a,b$ such that $h_{j_a} \neq h_{j_b}$, and therefore there exists an index $q \neq 0$ such that $\wh{h}^{(q)}_{j_1\dots j_N} \neq 0$. Hence, for any $s$ we have
\begin{equation}
  \Pc_s[\phi_{j_1} \otimes \dots \otimes \phi_{j_N}] \propto
   L_0^{(q)}  \cdot \Pc_{s-q}[\phi_{j_1} \otimes \dots \otimes \phi_{j_N}] 
\end{equation}
and therefore 
\begin{equation}
  \Pc_{-r-\rb}[\phi_{j_1} \otimes \dots \otimes \phi_{j_N}] \propto \left( L_0^{(q)} \right)^{[[-rq^{-1}]]} \left( \overline{L}_0^{(q)} \right)^{[[-\rb q^{-1}]]} \Pc_0 (\phi_{j_1} \otimes \dots \otimes \phi_{j_N})
\end{equation}
Thus, we can write $\Psi$ as $A_n \oplus \Ab_N$ descendant of $\Phi_{[j_1\dots j_N]}$, namely : 
\begin{align}
  \Psi \propto  \underbrace{L_{m_1}^{(r_1)} \dots L_{m_p}^{(r_p)}\left( L_0^{(q)} \right)^{[[-rq^{-1}]]}}_{\in A_N}  \cdot \underbrace{\Lb_{\mb_1}^{(\rb_1)} \dots \Lb_{\mb_\pb}^{(\rb_\pb)}
\left( \overline{L}_0^{(q)} \right)^{[[-\rb q^{-1}]]}}_{\in \Ab_N} \, \Phi_{[j_1\dots j_N]}
\end{align}
It follows that any invariant operator in the untwisted, non-diagonal sector is an $(A_N \oplus \Ab_N)$-descendant of some primary operator of the form $\Phi_{[j_1 \dots j_N]}$.

\paragraph{Diagonal untwisted operators.}
Let us first prove that the operators $\Phi_j^{(r)}$ are $A_N$ primary. 
When $r=0$, this follows from the fact that $\Phi_j^{(0)}$ is a highest weight-state of the full orbifold algebra. When $r \neq 0$ and $j \neq 1$, we consider
\begin{equation} \label{eq:A.Phi_j(r)}
\Psi = L_{m_1}^{(s_1)} \dots L_{m_p}^{(s_p)} \cdot L_{-1}^{(r)} \Lb_{-1}^{(-r)} \cdot \Phi_j^{(0)} \,,
  \quad \text{with} \ \begin{cases}
  m_1 \leq \dots \leq m_p \,, \\
  m=m_1+\dots+m_p>0\,, \\
  s_1+\dots+s_p = 0 \mod N \,.
  \end{cases}
\end{equation}
Furthermore $m_1, \dots, m_p \in \Zbb$ because $\Phi_j^{(0)}$ is untwisted.  But since $\Phi_j^{(0)}$ is a highest weight-state of the full orbifold algebra, $\Psi$ trivially vanishes unless  $m_i+\dots+m_p\leq 1$ for all $i$. But since  $m_1\leq \dots \leq m_p$ and $m \geq 1$, the only case to be considered is $m_p=1$ and $m_1=\dots=m_{p-1}=0$. So all we need to prove is that 
\begin{equation}
  L_0^{(s_1)} \dots L_0^{(s_{p-1})} \cdot L_1^{(s_p)} L_{-1}^{(r)} \Lb_{-1}^{(-r)} \cdot \Phi_j^{(0)}= 0 \, .
\end{equation}
We first rewrite the \emph{l.h.s.} using the orbifold Virasoro commutation rule
\begin{equation}
 \left[ L^{(r)}_1, L^{(s)}_{-1} \right] =2 L_{0}^{(r+s)}
\end{equation}
yielding
\begin{equation}
  L_0^{(s_1)} \dots L_0^{(s_{p-1})} \cdot L_1^{(s_p)} L_{-1}^{(r)} \Lb_{-1}^{(-r)} \cdot \Phi_j^{(0)}=
  2 \Lb_{-1}^{(-r)} \cdot L_0^{(s_1)} \dots L_0^{(s_{p-1})} \cdot L_0^{(s_p+r)}  \cdot \Phi_j^{(0)} \,.
\end{equation}
Using $L_0^{(s)}\cdot \Phi_j^{(0)} = Nh_j \delta_{s0} \Phi_j^{(0)}$, we get
\begin{equation}
   L_0^{(s_1)} \dots L_0^{(s_{p-1})} \cdot L_1^{(s_p)} L_{-1}^{(r)} \Lb_{-1}^{(-r)} \cdot \Phi_j^{(0)} =  2 (Nh_j)^p \, \delta_{s_1,0} \dots \delta_{s_{p-1},0} \cdot \delta_{s_p+r,0} \Lb_{-1}^{(-r)} \cdot \Phi_j^{(0)}  =0 
\end{equation}
since $s_1+\dots+s_N=0$ and $r \neq 0$. This shows that \eqref{eq:A.Phi_j(r)} must vanish, and hence $\Phi_j^{(r)}$ is primary under $A_N$. The case $r\neq 0$ and $j=1$ can be treated similarly: the condition on indices becomes $m_1=\dots=m_{p-1}=0$ and $m_p=2$, and then the relation $L_2^{(s)}L_{-2}^{(r)}\cdot\id=\frac{1}{2}Nc \delta_{r+s,0}$ can be used. 
\bigskip

Let's now consider a generic state of the form \eqref{eq_untwisted_generator_2} in the diagonal sector, namely 
\begin{equation}
  \Psi = \delta_{r+\rb,0} \,
  L_{m_1}^{(r_1)} \dots L_{m_p}^{(r_p)} \cdot \Lb_{\mb_1}^{(\rb_1)} \dots \Lb_{\mb_\pb}^{(\rb_\pb)}
  \cdot (\phi_j \otimes \dots \otimes \phi_j) \,,  \quad \text{with} \ \begin{cases}
  m_1 \leq \dots \leq m_p \leq -1 \,, \\
    \mb_1 \leq \dots \leq \mb_\pb \leq -1 \,.
  \end{cases}
\end{equation}
where  as above $r=r_1+\dots+r_p$ and $\rb=r_1+\dots+\rb_\pb$. 
 If $r=\rb=0$, then $\Psi$ is immediately  an $(A_N \oplus \Ab_N)$-descendant of $\Phi_j^{(0)}$. If $r=-\rb\neq 0$ and $j \neq 1$, we use
\begin{equation}
  L_1^{(-r)}L_{-1}^{(r)} \cdot (\phi_j \otimes \dots \otimes \phi_j)
  = \Lb_1^{(-\rb)} \Lb_{-1}^{(\rb)} \cdot (\phi_j \otimes \dots \otimes \phi_j)
  = 2Nh_j \, (\phi_j \otimes \dots \otimes \phi_j) \,,
\end{equation}
and rewrite $\Psi$ as
\begin{equation}
  \Psi = (2Nh_j)^{-2} \,
  \left( \underbrace{L_{m_1}^{(r_1)} \dots L_{m_p}^{(r_p)}  L_1^{(-r)}}_{\in A_N} \cdot \underbrace{ \Lb_{\mb_1}^{(\rb_1)} \dots \Lb_{\mb_\pb}^{(\rb_\pb)}
  \Lb_1^{(r)} }_{\in \Ab_N}\right) \cdot L_{-1}^{(r)}\Lb_{-1}^{(-r)} \cdot  (\phi_j \otimes \dots \otimes \phi_j) \,.
\end{equation}

In the case $r=-\rb\neq 0$ and $j=1$, we can proceed similarly and write
\begin{equation}
  \Psi =(Nc/2)^{-2} \,
  \left( L_{m_1}^{(r_1)} \dots L_{m_p}^{(r_p)} \cdot \Lb_{\mb_1}^{(\rb_1)} \dots \Lb_{\mb_\pb}^{(\rb_\pb)}
  \cdot L_2^{(-r)}\Lb_2^{(r)} \right) \cdot L_{-2}^{(r)}\Lb_{-2}^{(-r)} \cdot \id \,.
\end{equation}
This shows that $\Psi$ is an $(A_N \oplus \Ab_N)$-descendant of $\Phi_j^{(r)}$ or $\id^{(r)}$, respectively.
\bigskip

\paragraph{Twist operators.}
Let us prove that the operator $\sigma_j^{[k](r)}$ is primary under $A_N$. For $r=0$, this is straightforward, because by construction, the composite twist operator $\sigma_j^{[k]}$ is an $\OVir_N$ primary operator of conformal dimension $h_{\sigma_j}$. For $r\neq 0$, if $j \neq 1$ or $kr \neq 1 \mod N$, we consider
\begin{equation}  \label{eq:A.sigma_j}
  L_{m_1}^{(s_1)} \dots L_{m_p}^{(s_p)} \cdot L_{-\br{kr}/N}^{(r)} \Lb_{-\br{kr}/N}^{(-r)} \cdot \sigma_j^{[k]} \,,
  \quad \text{with} \ \begin{cases}
  m_1 \leq \dots \leq m_p \,, \\
  m=m_1+\dots+m_p>0 \,, \\
  s_1+\dots+s_p = 0 \mod N \,.
  \end{cases}
\end{equation}
If this quantity is non-zero, it is an $L_0^{(0)}$ eigenvector of eigenvalue $(h_{\sigma_j}+\br{kr}/N-m)$. Because of the condition $m_i \in -ks_i/N + \Zbb$, we have $m \in \Zbb$, and hence $m \geq 1$, so the eigenvalue is strictly smaller than $h_{\sigma_j}$, which is not possible within the module of $\sigma_j^{[k]}$. For $r=k^{-1}$ and $j=1$, we consider
\begin{equation}  \label{eq:A.sigma_id}
  L_{m_1}^{(s_1)} \dots L_{m_p}^{(s_p)} \cdot L_{-(N+1)/N}^{(k^{-1})} \Lb_{-(N+1)/N}^{(-k^{-1})} \cdot \sigma^{[k]}_\id \,,
  \quad \text{with} \ \begin{cases}
  m_1 \leq \dots \leq m_p \,, \\
  m=m_1+\dots+m_p>0 \,, \\
  s_1+\dots+s_p = 0 \mod N \,.
  \end{cases}
\end{equation}
If this quantity is non-zero, it is an $L_0^{(0)}$ eigenvector of eigenvalue $(h_{\sigma_j}-m+1+1/N)$. Using a similar argument to the one in the previous case, we find that $m$ is again a strictly positive integer: this gives a non-admissible eigenvalue, due to the null-vector relation $L_{-1/N}^{(k^{-1})} \cdot \sigma^{[k]}_\id=0$. Hence, \eqref{eq:A.sigma_j} and \eqref{eq:A.sigma_id} must vanish, and thus $\sigma_j^{[k](r)}$ and $\sigma_\id^{[k](r)}$ are primary under $A_N$. Similar proofs hold for $\Ab_N$.
\bigskip

Let us consider a generic invariant descendant operator in the twisted sector
\begin{equation}
  \Psi = \Pc_0\left[
    L_{m_1}^{(r_1)} \dots L_{m_p}^{(r_p)} \cdot \Lb_{\mb_1}^{(\rb_1)} \dots \Lb_{\mb_\pb}^{(\rb_\pb)}
    \cdot \sigma_j^{[k]} 
    \right] \,,
\end{equation}
where $\sigma_j^{[k]}$ is a composite twist operator associated to the primary operator $\phi_j$ of the mother CFT. By the commutation rules of $\Pi$ with the orbifold Virasoro modes, we have
\begin{equation}
  \Psi = \delta_{r+\rb,0} \cdot
    L_{m_1}^{(r_1)} \dots L_{m_p}^{(r_p)} \cdot \Lb_{\mb_1}^{(\rb_1)} \dots \Lb_{\mb_\pb}^{(\rb_\pb)}
    \cdot \sigma_j^{[k]} \,,
\end{equation}
where $r=r_1+\dots+r_p$ and $\rb=\rb_1+\dots+\rb_\pb$. If $r=\rb=0$, then $\Psi$ is immediately an $(A_N \oplus \Ab_N)$-descendant of $\sigma_j^{[k](0)}=\sigma_j^{[k]}$. For $r=-\rb \neq 0$, if $j \neq 1$ or $r \neq k^{-1}$, we use
\begin{equation}
  L_{\br{kr}/N}^{(-r)} L_{-\br{kr}/N}^{(r)} \cdot \sigma_j^{[k]} =
  \Lb_{\br{kr}/N}^{(r)} \Lb_{-\br{kr}/N}^{(-r)} \cdot \sigma_j^{[k]} = 
  \frac{2\br{kr} h_{\sigma_j}}{N} \, \sigma_j^{[k]} \,,
\end{equation}
and rewrite $\Psi$ as
\begin{equation}
  \Psi = 
  \left(\frac{N}{2\br{kr} h_{\sigma_j}} \right)^2
  \, \left(L_{m_1}^{(r_1)} \dots L_{m_p}^{(r_p)} \cdot \Lb_{\mb_1}^{(\rb_1)} \dots \Lb_{\mb_\pb}^{(\rb_\pb)}
  \cdot L_{\br{kr}/N}^{(-r)} \Lb_{\br{kr}/N}^{(r)}\right) \cdot L_{-\br{kr}/N}^{(r)} \Lb_{-\br{kr}/N}^{(-r)} \cdot\sigma_j^{[k]} \,.
\end{equation}
Similarly, if $j=1$ and $r=k^{-1}$, we have
\begin{equation}
  \Psi \propto
  \left(L_{m_1}^{(r_1)} \dots L_{m_p}^{(r_p)} \cdot \Lb_{\mb_1}^{(\rb_1)} \dots \Lb_{\mb_\pb}^{(\rb_\pb)}
  \cdot L_{(N+1)/N}^{(-k^{-1})} \Lb_{(N+1)/N}^{(k^{-1})} \right) \cdot L_{-(N+1)/N}^{(k^{-1})} \Lb_{-(N+1)/N}^{(-k^{-1})} \cdot\sigma_\id^{[k]} \,.
\end{equation}
This shows that $\Psi$ is an $(A_N \oplus \Ab_N)$-descendant of $\sigma_j^{[k](r)}$ or $\sigma_\id^{[k](r)}$, respectively.

We turn to the two-point function of twist operators,
\begin{equation}
  \aver{\sigma_i^{[k](r)}(z,\zb) \sigma_j^{[\ell](s)}(w,\wb)}
  = \aver{\sigma_i^{[k](r)}| \sigma_j^{[\ell](s)}} \times |z-w|^{-2h_{\sigma_i^{[k](r)}}-2h_{\sigma_j^{[\ell](s)}}} \,.
\end{equation}
This vanishes if $i \neq j$, because the modules of $\phi_i$ and $\phi_j$ are decoupled in the mother CFT.
This function also vanishes if $k+\ell \neq 0 \mod N$, by symmetry, under the cyclic permutation of the copies.
For $i=j$ and $k+\ell=0$, the scalar product $\aver{\sigma_j^{[k](r)}|\sigma_j^{[-k](s)}}$ is easily computed using the orbifold Virasoro commutation rules, which gives
\begin{equation}
  \aver{\sigma_j^{[k](r)}|\sigma_j^{[-k](s)}} = \delta_{r+s,0} \,.
\end{equation}

\section{Three-point functions}
\label{sec:proof-3pt}

\subsection{Three-point functions of twist operators}
\label{sec:appendixTTT}

Let us first state a useful property of correlation functions, which actually derives from the orbifold Ward identity \eqref{eq:orbifold-Ward}.
\bigskip

\noindent\textbf{Property.} Let $\Oc_1,\Oc_2,\Oc_3$ be three operators of the orbifold CFT, with twist charges $k_1,k_2,k_3$, respectively, such that $k_1=k_2+k_3 \mod N$. For any $r\in \Zbb_N$ and $q_1,q_2,q_3$ such that $q_i\in -k_ir/N+\Zbb$, and $q_1=q_2+q_3+1$, we have
\begin{align} \label{eq:Ward}
  \sum_{n=0}^\infty \left[
  b_n \aver{(L_{-q_1+n}^{(r)}\Oc_1^\dag) \Oc_2^\nodag \Oc_3^\nodag}
  - a_n \aver{\Oc_1^\dag (L_{q_2+n}^{(r)}\Oc_2^\nodag) \Oc_3^\nodag}
  + e^{i\pi q_2} b_n \aver{\Oc_1^\dag \Oc_2^\nodag (L_{q_3+n}^{(r)}\Oc_3^\nodag)}
  \right] = 0 \,,
\end{align}
where the coefficients $a_n,b_n$ are given by
\begin{equation}
  a_n = \frac{(q_3+1)q_3 \dots (q_3-n+2)}{n!} \,,
  \qquad b_n = (-1)^n \, \frac{(q_2+1)q_2 \dots (q_2-n+2)}{n!} \,.
\end{equation}
Note that, if $q_3=-1$ (resp. $q_2=-1$) then $a_n=\delta_{n0}$ (resp. $a_n=\delta_{n0}$). For any value of $q_1,q_2,q_3$, the sum in \eqref{eq:Ward} is finite, because, for any $\Oc_i$, there exists an integer $n_0$ such that $L_{\pm q_i+n}\cdot \Oc_i=0$ when $n \geq n_0$.

\noindent\textbf{Proof.}
Consider the integral,
\begin{equation}
  \oint dz \, (z-1)^{q_2+1} \, z^{q_3+1} \aver{\Oc_1^\dag(\infty) T^{(r)}(z) \Oc_2^\nodag(1) \Oc_3^\nodag(0)} \,,
\end{equation}
where the contour encloses the points $z=0$ and $z=1$. Since the prefactor $(z-1)^{q_2+1} \, z^{q_3+1}$ exactly compensates the monodromy of $T^{(r)}(z)$ around each of the $\Oc_i$'s, the integrand is single-valued, and hence the contour integral is closed. Using the Cauchy theorem, we can split the latter into a contour enclosing only $z=0$, and a contour enclosing only $z=1$. The Taylor expansions
\begin{equation}
  (1+u)^{q_3+1} = \sum_{n=0}^\infty a_n u^n \,,
  \qquad (1-u)^{q_2+1} = \sum_{n=0}^\infty b_n u^n \,,
  \qquad \text{for } |u|<1 \,,
\end{equation}
allow us to write the integrand for each contour as a power series in $z$ or $(z-1)$, which in turn yields \eqref{eq:Ward}.
$\blacksquare$
\bigskip

The goal of this section is to describe a recursive algorithm for the computation of OPE coefficients of the form
\begin{equation}
  \aver{\sigma_k^{\dag (-t)} \, \sigma_i^{(r)} \, (L_{-m_1}^{(r_1)} \dots L_{-m_p}^{(r_p)} \cdot \Phi_j^{(s)})} \,,
\end{equation}
where $p\geq 1$, and $m_1, \dots,m_p$ are integers, with $m_1 \geq \dots \geq m_p \geq 1$, and $r_1+\dots+r_p=0 \mod N$,
in terms of the quantities
\begin{equation} \label{eq:<s.s.L.Phi2>}
  \aver{\sigma_k \, \sigma_i \, (L_{-1}^{(u_1)} \dots L_{-1}^{(u_q)}
    \cdot \Lb_{-1}^{(\ub_1)} \dots \Lb_{-1}^{(\ub_\qb)} \cdot \Phi_j)} \,,
\end{equation}
where $u_1+\dots+u_q=-(\ub_1+\dots+\ub_\qb)=r+s-t \mod N$.
\bigskip

We start with the case $r=s=t=0$, namely
\begin{equation} \label{eq:<s.s.L.Phi>}
  \aver{\sigma_k^\dag \sigma_i^\nodag (L_{-m_1}^{(r_1)} \dots L_{-m_p}^{(r_p)} \cdot \Phi_j)} \,,
\end{equation}
Let us discuss the various cases, depending on the generator $L_{-m_1}^{(r_1)}$.
\begin{enumerate}

\item If $m_1=1$, then we have $m_1=\dots=m_p=1$, and the correlator \eqref{eq:<s.s.L.Phi>} is already of the form \eqref{eq:<s.s.L.Phi2>}.

\item If $r_1=0$, then the action of $L_{-m_1}^{(0)}$ inside a correlation function is easily expressed.
Using the commutator between $L_{-m_1}^{(0)}$ and a primary operator, we get
\begin{equation}
  \aver{\Phi_1 \Phi_2 (L_{-m_1}^{(0)}\cdot \Oc_3)}
  = (m_1h_2-h_1+h_3)\aver{\Phi_1 \Phi_2 \Oc_3} \,,
\end{equation}
where $\Phi_1,\Phi_2$ are primary, and $\Oc_3$ is any scaling operator. Here, $h_1,h_2,h_3$ are the conformal dimensions of $\Phi_1,\Phi_2,\Oc_3$, respectively.
Hence, we have
\begin{align}
  &\aver{\sigma_k^\dag \sigma_i^\nodag
    (L_{-m_1}^{(0)}\cdot L_{-m_2}^{(r_2)}  \dots L_{-m_p}^{(r_p)} \cdot \Phi_j)} \nn \\
  &\qquad = (m_1h_{\sigma_i}-h_{\sigma_k}+h_{\Phi_j}+m')
  \, \aver{\sigma_k^\dag \sigma_i^\nodag (L_{-m_2}^{(r_2)} \dots L_{-m_p}^{(r_p)} \cdot \Phi_j)} \,,
\end{align}
where $m'=m_2+\dots+m_p$. Thus, the problem of computing \eqref{eq:<s.s.L.Phi>} with the insertion of a chain of $p$ orbifold Virasoro generators, where the first one is of the form $L_{-m_1}^{(0)}$, has been reduced to the one with $(p-1)$ generators.

\item If $m_1>1$ and $r_1 \neq 0$, we shall use the linear relations \eqref{eq:Ward} as follows.

  First, we introduce a useful definition: for any operator $\Oc$, and $p \in \Nbb$, let $A_p(\Oc)$ be the space of descendants defined as
  \begin{equation}
    A_p(\Oc) = \mathrm{span}\left[ L_{-n_1}^{(u_1)} \dots L_{-n_q}^{(u_q)} \cdot \Oc \,,
      \quad q\leq p \,, \quad n_1 \geq \dots \geq n_q \geq 1 \}
      \right] \,.
  \end{equation}
  For instance, the operator $L_{-m_1}^{(r_1)}  \dots L_{-m_p}^{(r_p)} \cdot \Phi_j$ in \eqref{eq:<s.s.L.Phi>} belongs to $A_p(\Phi_j)$.

  By convention, we take $1 \leq r \leq N-1$. For any integer $m'_1 \in \{2,3,\dots,m_1\}$, if we set $q_1=2-r_1/N-m'_1,q_2=1-r_1/N,q_3=-m'_1$ in \eqref{eq:Ward}, we get a linear relation of the form
\begin{equation} \label{eq:Ward2}
  \sum_{n=0}^{m'_1+(m_2+\dots+m_p)} b_n \, \aver{\sigma_k^\dag \sigma_i^\nodag (L_{-m_1'+n}^{(r_1)}
    \cdot L_{-m_2}^{(r_2)} \dots L_{-m_p}^{(r_p)}\cdot \Phi_j)} = 0 \,.
\end{equation}
We rewrite this as
\begin{equation} \label{eq:Ward3}
  \sum_{n=2}^{m'_1} b_{m'_1-n} \, \aver{\sigma_k^\dag \sigma_i^\nodag (L_{-n}^{(r_1)}
    \cdot L_{-m_2}^{(r_2)} \dots L_{-m_p}^{(r_p)}\cdot \Phi_j)} = B_{m'_1} \,,
\end{equation}
where the right-hand side
\begin{equation}
  B_{m'_1}  = -\sum_{n=-1}^{m_2+\dots+m_p} b_{m'_1+n} \, \aver{\sigma_k^\dag \sigma_i^\nodag (L_n^{(r_1)}
    \cdot L_{-m_2}^{(r_2)} \dots L_{-m_p}^{(r_p)}\cdot \Phi_k)} \,,
  \end{equation}
can be treated with the $\OVir_N$ commutation relations, to give a linear combination of the form
\begin{equation}
  B_{m'_1} \in -b_{m'_1-1} \aver{\sigma_k^\dag \sigma_i^\nodag
    (L_{-m_2}^{(r_2)} \dots L_{-m_p}^{(r_p)}\cdot L_{-1}^{(r_1)}\cdot \Phi_j)} + \aver{\sigma_k^\dag \sigma_i^\nodag \, A_{p-1}(\Phi_j)} \,.
\end{equation}
The relations \eqref{eq:Ward3} for $m'_1=2,\dots,m_1$ form an invertible triangular $(m_1-1) \times (m_1-1)$ linear system for the coefficients $\aver{\sigma_k^\dag \sigma_i^\nodag (L_{-n}^{(r_1)} \cdot L_{-m_2}^{(r_2)} \dots L_{-m_p}^{(r_p)}\cdot \Phi_j)}$ with $n=2,\dots,m_1$. By solving this system, on gets \eqref{eq:<s.s.L.Phi>} in terms of the $B_{m'_1}$'s. Thus, we have reduced the computation of \eqref{eq:<s.s.L.Phi>} to that of OPE coefficients of the form
\begin{equation}
  \aver{\sigma_k^\dag \sigma_i^\nodag (L_{-n_1}^{(u_1)} \dots L_{-n_q}^{(u_q)} \cdot \Phi_j)}
  \quad \text{and} \quad
  \aver{\sigma_k^\dag \sigma_i^\nodag (L_{-n_1}^{(u_1)} \dots L_{-n_q}^{(u_q)}\cdot L_{-1}^{(r_1)}\cdot \Phi_j)} \,,
\end{equation}
with $0 \leq q \leq p-1$.

\end{enumerate}

Hence, in the three cases, the above steps define an algorithm to compute \eqref{eq:<s.s.L.Phi>} in terms of \eqref{eq:<s.s.L.Phi2>}, by recursion on $p$.
\bigskip

We now turn to OPE coefficients of the form
\begin{equation} \label{eq:<s^t.s^r.Phi^s>}
  \aver{\sigma_k^{\dag (-t)} \sigma_i^{(r)} \Phi_j^{(s)}} \,,
\end{equation}
with generic values of the Fourier indices $r,s,t \in \Zbb_N$. By convention, we take $0 \leq r,s,t \leq N-1$, and we introduce the notations $(\bar{r},\bar{s},\bar{t})=(N-r,N-s,N-t)$.
Recall the definitions
\begin{equation}
  \begin{aligned}
    & \sigma_i^{(r)} := \mathrm{const} \times L_{-r/N}^{(r)}\Lb_{-r/N}^{(\rb)} \cdot \sigma_i \,, \\
    & \Phi_j^{(s)} := \mathrm{const} \times L_{-1}^{(s)}\Lb_{-1}^{(\sbar)} \cdot \Phi_j \,, \\
    & \sigma_k^{\dag (-t)} := \mathrm{const} \times L_{-t/N}^{(\tb)}\Lb_{-t/N}^{(t)} \cdot \sigma_k^\dag \,,
    \end{aligned}
\end{equation}
for non-zero $r,s,t$. Here, we have assumed that $\phi_i,\phi_j,\phi_k$ is different from $\id$, but the argument is easily adapted otherwise.

Using the Ward identity \eqref{eq:Ward} with $q_1=q_2=t/N,q_3=-1$, for the insertion of $T^{(\tb)}(z)$ in the function $\aver{(\Lb_{-t/N}^{(t)}\sigma_k^\dag) \, \sigma_i^{(r)} \, \Phi_j^{(s)}}$, we get
\begin{align}
  & \mathrm{const} \times \aver{\sigma_k^{\dag (-t)} \sigma_i^{(r)} \Phi_j^{(s)}} \nn \\
  & = \aver{(\Lb_{-t/N}^{(t)}\sigma_k^\dag) (L_{t/N}^{(\tb)}\sigma_i^{(r)}) \Phi_j^{(s)}}
  - e^{i\pi \rb/N} \sum_{n=0}^{2} b_n \aver{(\Lb_{-t/N}^{(t)}\sigma_k^\dag) \sigma_i^{(r)} (L_{n-1}^{(\tb)}\Phi_j^{(s)})}  \,. \label{eq:Ward4}
\end{align}
If $t>r$, then $L_{t/N}^{(\tb)}\sigma_i^{(r)}=0$, and the first term on the right-hand side vanishes. If $t \leq r$, using the commutation relations, we write this first term as
\begin{equation}
  \mathrm{const} \times \frac{t+r}{N} \aver{(\Lb_{-t/N}^{(t)}\sigma_k^\dag) (L_{(t-r)/N}^{(r-t)}\Lb_{-r/N}^{(-r)}\sigma_i^\nodag) \Phi_j^{(s)}} \,,
\end{equation}
and then we use again the identity \eqref{eq:Ward} with $q_1=q_2=(t-r)/N,q_3=-1$ for the insertion of $T^{(r-t)}(z)$ in $\aver{(\Lb_{-t/N}^{(t)}\sigma_k^\dag) (\Lb_{-r/N}^{(-r)}\sigma_i^\nodag) \Phi_j^{(s)}}$, which yields
\begin{align}
  & \aver{(\Lb_{-t/N}^{(t)}\sigma_k^\dag) (L_{(t-r)/N}^{(r-t)}\Lb_{-r/N}^{(-r)}\sigma_i^\nodag) \Phi_j^{(s)}}
    = \delta_{rt} h_{\sigma_k} \aver{(\Lb_{-t/N}^{(t)}\sigma_k^\dag) (\Lb_{-r/N}^{(-r)}\sigma_i^\nodag) \Phi_j^{(s)}} \nn \\
  & \qquad - e^{i\pi (t-r)/N} \sum_{n=0}^{2} b'_n \aver{(\Lb_{-t/N}^{(t)}\sigma_k^\dag) (\Lb_{-r/N}^{(-r)}\sigma_i^\nodag) (L_{n-1}^{(r-t)} \Phi_j^{(s)})}  \,.
\end{align}
For the second term in \eqref{eq:Ward4}, if $r\neq 0$, we have
\begin{equation}
  \aver{(\Lb_{-t/N}^{(t)}\sigma_k^\dag) \sigma_i^{(r)} (L_{n-1}^{(\tb)}\Phi_j^{(s)})}
  = \mathrm{const} \times \aver{(\Lb_{-t/N}^{(t)}\sigma_k^\dag) (L_{-r/N}^{(r)}\Lb_{-r/N}^{(\rb)}\sigma_i) (L_{n-1}^{(\tb)}\Phi_j^{(s)})} \,,
\end{equation}
which, upon applying \eqref{eq:Ward}, yields 
 \begin{equation}
  \mathrm{const} \times \sum_{p=0}^{2-n} b''_p \aver{(\Lb_{-t/N}^{(t)}\sigma_k^\dag) (\Lb_{-r/N}^{(\rb)}\sigma_i^\nodag)
    (L_{p-1}^{(r)}L_{n-1}^{(\tb)}\Phi_j^{(s)})} \,.
\end{equation}

Hence, we have shown how to express \eqref{eq:<s^t.s^r.Phi^s>} in the form
\begin{equation}
  \aver{\sigma_k^{\dag (-t)} \sigma_i^{(r)} \Phi_j^{(s)}}
  = \aver{(\Lb_{-t/N}^{(t)}\sigma_k^\dag)(\Lb_{-r/N}^{(\rb)}\sigma_i^\nodag) (\lambda \cdot \Phi_j)} \,,
\end{equation}
where $\lambda$ is a linear combination of generators of the form $L_{-m_1}^{(r_1)}\dots L_{-m_p}^{(r_p)}$ with $r_1+\dots+r_p= r+s-t \mod N$. In other words, we have ``pushed'' the orbifold Virasoro generators entering the definition of $\sigma_k^{\dag (-t)}$ and $\sigma_i^{(r)}$, to translate them into an action of $\OVir_N$ on $\Phi_j^{(s)}$. Proceeding similarly with the $\overline{\OVir}_N$ modes, we express \eqref{eq:<s^t.s^r.Phi^s>} as
\begin{equation}
  \aver{\sigma_k^{\dag (-t)} \, \sigma_i^{(r)} \, \Phi_j^{(s)}}
  = \aver{\sigma_k^\dag \, \sigma_i^\nodag \, (\lambda \cdot \bar\lambda \cdot \Phi_j)} \,,
\end{equation}
where $\lambda$ is the same as above, and $\bar\lambda$ is obtained from $\mu$ by the change $L_{-m_j}^{(r_j)} \to \Lb_{-m_j}^{(-r_j)}$. We can then use the algorithm described above in the case $r=s=t=0$, to express $\aver{\sigma_k^{\dag (-t)} \, \sigma_i^{(r)} \, \Phi_j^{(s)}}$ in terms of \eqref{eq:<s.s.L.Phi2>}. Finally, extending this line of argument for $\aver{\sigma_k^{\dag (-t)} \, \sigma_i^{(r)} \, \Phi_j^{(s)}}$ to the general case of \eqref{eq:<s.s.L.Phi>} is straightforward.

\subsection{Calculation of a three-point function in the $\Zbb_3$ orbifold of the Yang-Lee CFT}
\label{sec:appendixunfoldYLZ3}

A useful technique , which is often employed in the literature \cite{calabrese_entanglement_2009,calabrese_entanglement_2004,cardy_entanglement_2016} is to \textit{unfold} the three-point function to a mother CFT correlator defined on $\mathbb{C}$. Let us consider, as a simple example, the three-point function:
\begin{equation}
  \mathcal{C}^{[\id,\phi,\phi]}_{\sigma_\phi^\nodag,\sigma_\phi^\dag}
  =\aver{\sigma_\phi \cdot [\id,\phi,\phi] \cdot \sigma_\phi^\dag} \,,
\end{equation}
in the $\Zbb_3$ orbifold. This can be expressed as a correlator on the replicated Riemann surface $\Sigma_3$ (with $\Sigma_3$ conformally equivalent to the Riemann sphere $\mathbb{C}$):
\begin{equation}
  \mathcal{C}^{[\id,\phi,\phi]}_{\sigma_\phi^\nodag,\sigma_\phi^\dag}
  = \sqrt{3} \, \aver{\phi(\infty,\infty) \phi(1,1) \phi(e^{2\pi i},e^{2\pi i}) \phi(0,0)}_{\Sigma_3} \,,
\end{equation}
which we map to $\mathbb{C}$, through $z\mapsto w=z^{1/3}$, to find:
\begin{equation}
  \mathcal{C}^{[\id,\phi,\phi]}_{\sigma_\phi^\nodag,\sigma_\phi^\dag} = \frac{\sqrt{3}}{3^{4 h_{\phi}}}
  \aver{\phi(\infty,\infty) \phi(1,1)\phi(e^{2\pi i/3},e^{-2\pi i/3}) \phi(0,0)}_{\Cbb} \,.
\end{equation}
Now, we use the result of \cite{cardy_conformal_1985} to express the four-point function $\aver{\phi | \phi(1,1)\phi(w,\bar{w})|\phi}$ in terms of hypergeometric functions:
\begin{align}
  & \aver{\phi | \phi(1,1)\phi(w,\bar{w})|\phi} \nn \\
  & \qquad = |w|^{4/5}|1-w|^{4/5} \left[
    \left|{}_2\mathrm{F}_1 \left(\left.\frac{3}{5},\frac{4}{5},\frac{6}{5} \right| w \right) \right|^2
    +(C_{\phi\phi}^{\phi})^2 \left|w^{-1/5} \, {}_2\mathrm{F}_1\left(\left.\frac{3}{5},\frac{2}{5},\frac{4}{5} \right| w \right) \right|^2 \right] \,,
\end{align}
with the Yang-Lee CFT structure constant given by:
\begin{equation}
  C_{\phi\phi}^\phi =
  \frac{i \sqrt{\frac{1}{2}(3 \sqrt{5}-5)} \, \Gamma\left(\frac{1}{5}\right)^{3}}{10 \pi \Gamma\left(\frac{3}{5}\right)} \,.
\end{equation}
Thus, we find:
\begin{equation} \label{eq:yangleeZ3C}
  \mathcal{C}^{[\id,\phi,\phi]}_{\sigma_\phi^\nodag,\sigma_\phi^\dag} = -11.054494 \neq 0 \,.
\end{equation}

\section{Proofs for modular matrices}
\label{sec:proof-modular}

\subsection{The modular matrices $P_n$}
Let us prove some useful properties of the matrices $P_n$'s defined in (\ref{eq:Pn}--\ref{eq:qn}):
\begin{align}
  & P_{n+N}=P_n \,, \label{eq:P_{n+N}} \\
  & P_n \cdot P_{-n^{-1}} = \id \,, \label{eq:Pinv} \\
  & P_n^t = P_{n^{-1}} \,, \label{eq:Pt} \\
  & \overline{P}_n = P_{-n} \,, \label{eq:Pbar} \\
  & P_n^\dag \, P_n^{\phantom{\dag}} = \id \,, \label{eq:Punitary} \\
  & T^{1/N} P_n \, T^{1/N} P_{1-n^{-1}} \, T^{1/N} = P_{n-1} \,, \quad \text{for } n \neq 1 \mod N \,. \label{eq:TPTPT}
\end{align}
The notations $P_{-n^{-1}},P_{n^{-1}},P_{1-n^{-1}}$, where $n^{-1}$ is the inverse of $n$ in $\Zbb_N^\times$, are justified by the property \eqref{eq:P_{n+N}}, which means that $P_n$ is actually defined for $n \in \Zbb_N^\times$.
\bigskip

\begin{itemize}
  
\item To prove \eqref{eq:P_{n+N}}, we remark that the integer $[[-n^{-1}]]$ is unchanged under $n \to n+N$, so we have
\begin{equation}
  P_{n+N} = T^{-n/N-1} \cdot Q_{n+N} \cdot T^{[[-n^{-1}]]/N} \,.
\end{equation}
On the other hand, from \eqref{eq:qn} we have $q_{n+N}(\tau) = q_n(\tau)+1$, and hence $Q_{n+N}=T\cdot Q_n$. As a result, we get $P_{n+N}=P_n$.

\item For \eqref{eq:Pinv}, we use the identity $q_n[q_{[[-n^{-1}]]}(\tau)]=\tau$, which yields, for $0<n<N$:
\begin{equation}
  P_n \cdot P_{-n^{-1}} = T^{-n/N} \cdot Q_n \cdot Q_{[[-n^{-1}]]} \cdot T^{n/N} =\id \,.
\end{equation}

\item To study the transpose $P_n^t$ and the conjugate $\overline{P}_n$, we use the following properties of the modular group. Since $S^2=\id$, any element of the group can be written in the form
  \begin{equation}
    T^{p_1} \cdot S \cdot T^{p_2} \dots S \cdot T^{p_k} \,,
  \end{equation}
  where $p_1,\dots,p_k \in \Zbb$.
  Let $t(\tau)=\tau+1$ and $s(\tau)=-1/\tau$.
  Let us show, by induction on $k$, that the following three equations are equivalent:
  \begin{align}
    & (t^{p_1} \circ s \circ t^{p_2} \dots s \circ t^{p_k})(\tau) = \frac{a\tau+b}{c\tau+d} \,, \label{eq:mod1} \\
    & (t^{p_k} \circ s \circ t^{p_{k-1}} \dots s \circ t^{p_1})(\tau) = \frac{d\tau+b}{c\tau+a} \,,  \label{eq:mod2} \\
    & (t^{-p_1} \circ s \circ t^{-p_2} \dots s \circ t^{-p_k})(\tau) = \frac{-a\tau+b}{c\tau-d} \,.  \label{eq:mod3}
  \end{align}
  The proof goes as follows. For $k=2$, we have
  \begin{equation}
    t^p \circ s \circ t^q = \frac{p \tau + (pq-1)}{\tau+q} \,,
  \end{equation}
  and hence the equivalence is straightforward. Now suppose that the above equivalence holds for some $k\geq 2$. We can write
  \begin{align}
    & (t^{p_1} \circ s \circ \dots s \circ t^{p_{k+1}})(\tau)
      = \frac{a (s \circ t^{p_{k+1}})(\tau)+b}{c(s \circ t^{p_{k+1}})(\tau)+d}
      = \frac{b\tau+(bp_{k+1}-a)}{d\tau+(dp_{k+1}-c)} \,, \\
    & (t^{p_{k+1}} \circ s \dots s \circ t^{p_1})(\tau)
      = (t^{p_{k+1}} \circ s)\left(\frac{d\tau+b}{c\tau+a}\right)
      = \frac{(dp_{k+1}-c)\tau+(bp_{k+1}-a)}{d\tau+b} \,, \\
    & (t^{-p_1} \circ s \circ t^{-p_2} \dots s \circ t^{-p_{k+1}})(\tau)
      = \frac{-a(s \circ t^{-p_{k+1}})(\tau)+b}{c(s \circ t^{-p_{k+1}})(\tau)-d}
      = \frac{-b\tau+(bp_{k+1}-a)}{d\tau-(dp_{k+1}-c)} \,,
  \end{align}
  and thus the equivalence holds also for $k+1$.
  An important consequence is that, if the matrix for \eqref{eq:mod1} is  $M$, then the matrices for \eqref{eq:mod2} and \eqref{eq:mod3} are given by $M^t$ and $\overline{M}$ respectively.

  We apply the equivalence to the matrices $Q_n$.
  Let us take $0<n<N$ for convenience, and denote $a_n=[[-n^{-1}]]$ and $b_n=(n[[-n^{-1}]]+1)/N$, so that
  \begin{equation}
    Nb_n -n a_n = 1 \,,
    \qquad  q_n(\tau) = \frac{n\tau-b_n}{N\tau-a_n} \,.
  \end{equation}
  
  Changing $n\mapsto -a_n$ gives $(a_n,b_n)\mapsto(N-n,b_n-a_n)$, whereas $n \mapsto -n$ gives $(a_n,b_n)\mapsto(N-a_n,b_n-n)$. Hence, we have
  \begin{equation}
    q_{-a_n}(\tau) = \frac{-a_n(\tau-1)-b_n}{N(\tau-1)+n} \,,
    \qquad q_{-n}(\tau) = \frac{-n(\tau-1)-b_n}{N(\tau-1)+a_n} \,,
  \end{equation}
  which yields $Q_{-a_n}=Q_n^t \cdot T^{-1}$ and $Q_{-n}=\overline{Q}_n \cdot T^{-1}$, and finally
  \begin{equation}
    P_{-a_n}=T^{a_n/N} \cdot Q_n^t T^{-n/N} = P_n^t \,,
    \qquad P_{-n}=T^{n/N} \cdot \overline{Q}_n T^{-a_n/N} = \overline{P}_n \,.
  \end{equation}

\item The combination of (\ref{eq:Pinv}--\ref{eq:Pbar}) yields the unitarity of $P_n$ \eqref{eq:Punitary}.

\item For $n \neq 1 \mod N$, we define $a_n=[[-n^{-1}]]$ and $a'_n=[[-(1+a_n)^{-1}]]$. There exist two integers $b_n,b'_n$ such that
  \begin{equation}
    Nb_n-na_n=1 \,, \qquad Nb'_n-(1+a_n)a'_n=1 \,.
  \end{equation}
  A simple calculation gives
  \begin{equation}
    q_n[q_{1+a_n}(\tau)] = \frac{(n-1)\tau-b_n''}{N\tau-a''_n} \,,
  \end{equation}
  where $a''_n=Nb'_n-a_na'_n$ and $b''_n=nb'_n-b_na'_n$, and hence $Nb''_n-(n-1)a''_n=1$. Hence, if $0<a''_n<N$ we have $Q_nQ_{1+a_n}=Q_{n-1}$ (otherwise we can always shift $a''_n$ by a multiple of $N$). This yields the identity \eqref{eq:TPTPT}.
\end{itemize}

For $n=1$, we have $[[-1]]=N-1$, and
\begin{equation}
  q_1(\tau) = \frac{\tau-1}{N(\tau-1)+1} = (s \circ t^{-N} \circ s \circ t^{-1})(\tau) \,,
\end{equation}
which yields
\begin{equation} \label{eq:P1}
  P_1 = T^{-1/N} S T^{-N} S T^{-1/N} \,,
  \qquad
  P_{-1} = T^{1/N} S T^{N} S T^{1/N} \,.
\end{equation}

\subsubsection*{A recursion relation for the matrices $P_n$}
In this section, we shall reproduce and prove a recursion relation for the $P_n$ matrices found in \cite{bantay_permutation_2002}, which provides their decomposition into $S$ and $T$ matrices after a relatively small number of recursive steps.

The key idea is to relate  $P_n$ matrices defined at \textit{different N}. We will first make the dependence on $N$ of the $P_n$ and $Q_{n}$ matrices explicit through the notation:
\begin{equation}
    P_n\rightarrow P_{n|N} \quad   Q_{n}\rightarrow Q_{n|N}
\end{equation}
We now consider the product of matrices
\begin{equation}
    T^{N/n}ST^{n/N}P_{n|N}T^{1/(nN)}
\end{equation}
We substitute in the above the definition of (\ref{eq:Pn}) in the above to find:
\begin{equation}
     T^{N/n}S\,T^{n/N}P_{n|N}T^{1/(nN)}= T^{N/n}S\, Q_{n|N}T^{b_n/n}
\end{equation}
Now the $S$ matrix is given by:
\begin{equation}
    S=\left(\begin{array}{cc}
        0 & 1 \\
        -1 & 0
    \end{array}\right)
\end{equation}
while the $Q_{n|N}$ matrix is given by:
\begin{equation}
    Q_{n|N}= \left( \begin{array}{cc}
        n & -b_n  \\
        N & -a_n
    \end{array}\right)
\end{equation}
with $N b_n- n a_n=1$. Acting from the left on the above with the $S$ matrix, one finds:
\begin{equation}
    S\, Q_{n|N}= \left( \begin{array}{cc}
        N & -a_n  \\
        -n & b_n
    \end{array}\right)=Q_{N|-n}
\end{equation}
since 
$Nb_n-n a_n=1$ is equivalent to $(-n)a_n-b_n N=1$. Next, we have by the definition in (\ref{eq:Pn}) that:
\begin{equation}
    P_{N|-n}=T^{N/n} Q_{n|N} T^{a_n/n}
\end{equation}
so that we have effectively established a recursive relation:
\begin{equation}\label{eq:recursionacrossN}
   \boxed{    P_{n|N}= T^{-n/N}ST^{-N/n}P_{N|-n}T^{-1/(nN)}}
\end{equation}
for which the recursion ends with a term $P_{0|N}=P_{0|1}\equiv S$. 
\subsubsection*{Example for $N=5$}
Suppose we want to calculate the  $P_{2|5}$ matrix. We use the recursion relation to find:
\begin{equation}
    P_{2|5}=T^{-2/5}ST^{-5/2}P_{5|-2}T^{-1/10}
\end{equation}
Noting that $P_{5|-2}=P_{1|-2}$, and  using once more the recursion relation, we find:
\begin{equation}
    P_{1|-2}=T^{1/2}ST^{2}P_{-2|1}T^{1/2}
\end{equation}
But $P_{-2|1}=P_{0|1}=S$, so, one can put everything together to find:
\begin{equation}
      P_{2|5}=T^{-2/5}ST^{-2}ST^{2}ST^{2/5}
\end{equation}

\subsection{The orbifold modular $\mathcal{S}$-matrix}
In this section, we will show that the solution for the orbifold $\mathcal{S}$-matrix we have found satisfies the properties (\ref{eq:St=S})-(\ref{eq:(ST)^3=C}).  In our calculations, the  $\Sc$ and $\Tc$ matrices (and any products made with them), are evaluated ``block-by-block'', in the sense that for each proof we provide here, the external indices are restricted to correspond to only one type of operator -- non-diagonal (ND), diagonal(D), twisted (T). For example, checking unitarity in the D-D block means proving that:
\begin{equation}
    (\Sc\Sc^\dag)_{i^{(r)},j^{(s)}}= \delta_{i^{(r)},j^{(s)}} \,,
\end{equation}
for generic diagonal operator labels $i^{(r)},j^{(s)}$.

We will only present in this section some of the more technical demonstrations, since the rest can be quickly reproduced by the interested reader  through similar or simpler arguments.

\subsubsection{Symmetry of $\mathcal{S}$}
As a warm-up, let's prove the symmetry of the $\Sc$ matrix. The check is straightforward everywhere but in the TT block, where we need to employ the property (\ref{eq:Pt}) to find:
\begin{equation}
    \Sc^t_{i^{[k](r)}, j^{[\ell](s)}} = \frac{\omega^{-ks-\ell r} \, (P_{k.\ell^{-1}})_{ji}}{N}=\frac{\omega^{-ks-\ell r} \, (P_{\ell.k^{-1}})_{ij}}{N}=  \Sc_{i^{[k](r)}, j^{[\ell](s)}} \,.
\end{equation}

\subsubsection{Unitarity of $\Sc$}
We present in this section the more involved proofs for the unitarity of $\Sc$ in the DD and TT blocks.

\paragraph{In the DD block.}
We want to check the property:
\begin{equation} \label{eq:Dblocksconstraint}
  \sum_\alpha \mathcal{S}_{i^{(r)},\alpha} \mathcal{S}^{*}_{i'^{(r')},\alpha}=\delta_{(i,r),(i',r')} \,,
\end{equation}
where the sum in $\alpha$ runs over all the invariant primary operators in the orbifold.
We first consider the sum over ND operators :
\begin{equation}
  \sum_{J = [j_1,...,j_N]}\mathcal{S}_{(i,r),J} \mathcal{S}^{*}_{J,(i',r')} \,,
\end{equation}
which, using the known form of the S-matrix elements in this sector, is:
\begin{equation}
    \sum_{J} S_{i,j_1} \dots S_{i,j_N} \cdot S_{i',j_1}\dots S_{i',j_N} \,,
\end{equation}
where the sum is over equivalence classes $J$ of the $N$-tuples $(j_1,\cdots,j_N)$ under $\Zbb_N$, with the exception of $N$-tuples for which the indices are equal. This can be expressed as:
\begin{equation}
     \sum_{J = [j_1,...,j_N]} = \frac{1}{N}\left(\sum_{(j_1,\cdots,j_N)}-\sum_{(j,\cdots,j)}\right) \,,
\end{equation}
where the first sum on the RHS is over all $N$-tuples $(j_1,\dots j_N)$. This means each $j_i$ runs over the whole mother CFT spectrum so that the sum over the ND indices can be conveniently expressed as:
\begin{equation}
   \sum_{J}\mathcal{S}_{i^{(r)},J} \mathcal{S}^{*}_{J,i'^{(r')}}=\frac{\delta_{i,i'}}{N}-\frac{\sum_{(j,r)} S_{i,j}^N S_{i',j}^N}{N^2}  
\end{equation}
On the other hand, the sum over D operators is given by:
\begin{equation}
     \sum_{j,s}\mathcal{S}_{i^{(r)},j^{(s)}} \mathcal{S}^{*}_{j^{(s)},i'^{(r')}}=\frac{\sum_{j,r} S_{ij}^N S_{i'j}^N}{N^2} 
\end{equation}
so that we are just left with checking that the sum over T operators satisfies
\begin{equation}
     \sum_{j,s,k}\mathcal{S}_{i^{(r)},j^{[k](s)}} \mathcal{S}^{*}_{j^{[k](s)},i'^{(r')}}=\frac{N\delta_{r,r'}-1}{N}\delta_{i,i'}
\end{equation}
which is indeed the case.

\paragraph{In the TT block.}
We want to prove in this section that:
\begin{equation}
  \sum_\alpha \mathcal{S}_{i^{[k](r)},\alpha} \mathcal{S}^{*}_{\alpha,i'^{[k'](r')}}
  =\delta_{i,i'}\delta_{r,r'}\delta_{k,k'}
\end{equation}
where $\alpha$ runs once again over all invariant primary operators of the orbifold. 

To start off, the sum over non-diagonal entries vanishes, since the $\Sc$-matrix entries indexed by a twisted operator and a non-diagonal operator vanish. On the other hand, the sum over diagonal field entries is easily calculated to be:
\begin{equation}\label{eq:part1unitarityTT}
     \sum_{(j,s)}\mathcal{S}_{i^{[k](r)},(j,s)} \mathcal{S}^{*}_{(j,s),i'^{[k'](r')}}=\frac{1}{N^2} \sum_{(j,s)} w^{s(k-k')}S_{ij}S_{ji'}=\frac{\delta_{i,i'}\delta_{k,k'}}{N}
\end{equation}
Finally, we now consider the sum over T entries:
\begin{equation}
     \sum_{j^{[t](s)}}\mathcal{S}_{i^{[k](r)},j^{[t](s)}} \mathcal{S}^{*}_{j^{[t](s)},i'^{[k'](r')}}
\end{equation}
We substitute our results and find:
\begin{equation}
  \sum_{j,s,t} \mathcal{S}_{i^{[k](r)},j^{[t]s}}\, \mathcal{S}^*_{i'^{[k'](r')},j^{[t](s)}} 
  =\frac{1}{N^2}\sum_{j,s,t} \omega^{-s(k-k')-t(r-r')} (\mathbf{P}_{tk^{-1}})_{ij}(\mathbf{P}^{\dagger}_{tk'^{-1}})_{ji'} \,.
\end{equation}
We now sum over $s$, to obtain a factor $\delta_{k^{-1},k'^{-1}}=\delta_{k,k'}$ .
Then, we  perform the sum over $j$, and use the unitarity of the $P_n$ matrices (\ref{eq:Punitary}) to arrive at:
\begin{equation}
\sum_{j,s,t} \mathcal{S}_{i^{[k](r)},j^{[t]s}}\, \mathcal{S}^*_{i'^{[k'](r')},j^{[t](s)}} =\frac{1}{N}\sum_{t} \omega^{-t(r-r')}\delta_{i,i'}\delta_{k,k'} 
\end{equation}
Finally, we sum over $t$ and find:
\begin{equation}
    \sum_{j,s,t} \mathcal{S}_{i^{[k](r)},j^{[t]s}}\, \mathcal{S}^*_{i'^{[k'](r')},j^{[t](s)}}=\delta_{i,i'}\delta_{k,k'}\delta_{r,r'}-\frac{1}{N}\delta_{i,i'}\delta_{k,k'} \,,
\end{equation}
which, added to (\ref{eq:part1unitarityTT})  gives the sought after result.

\subsubsection{The relation $(\mathcal{ST})^3=\mathcal{C}$}

The constraint \eqref{eq:(ST)^3=C} on the modular data of the CFT is equivalent to:
\begin{equation}
  \mathcal{STS}=\mathcal{T}^{\dagger}\mathcal{S}\mathcal{T}^{\dagger}
\end{equation}
As in the previous section, we will show that the relation above holds in the DD and TT blocks.

\paragraph{In the DD block.}
We want to check the constraint:
\begin{equation}\label{eq:RHSofXDD}
   \sum_\alpha \mathcal{S}_{i^{(r)},\alpha}\mathcal{T}_{\alpha} \mathcal{S}_{\alpha,i'^{(r')}}=\mathcal{T}^{\dagger}_{i^{(r)}} \mathcal{S}_{i^{(r)},i'^{(r')}} \mathcal{T}^{\dagger}_{i'^{(r')}}
\end{equation}
where the sum $\alpha$ runs over all the invariant primary operator labels.

Substituting our expressions for the $\Sc$ and $\Tc$ matrices, we find that the sum over entries labelled by untwisted operators gives:
\begin{equation}
     \sum_{\text{untwisted } \alpha}\mathcal{S}_{i^{(r)},\alpha}\,\mathcal{T}_\alpha \, \mathcal{S}_{\alpha,i'^{(r')}}=\frac{1}{N}\left(STS\right)^N_{ii'}
\end{equation}
while the sum over twisted operators vanishes.
The RHS of (\ref{eq:RHSofXDD}) can also be conveniently written as:
\begin{equation}
   \mathcal{T}^{\dagger}_{i^{(r)}} \mathcal{S}_{i^{(r)},i'^{(r')}} \mathcal{T}^{\dagger}_{i'^{(r')}}=\frac{1}{N}\left(TST\right)_{i,i'}^N
\end{equation}
Since in the mother CFT the modular matrices are constrained by
\begin{equation}
    STS=T^{\dagger}S T^{\dagger}
\end{equation}
we find that the relation (\ref{eq:(ST)^3=C}) is indeed satisfied in this block.

\paragraph{In the TT block.}
We want to check the constraint:
\begin{equation}\label{eq:TTbllockST3}
   \sum_{\alpha\in \{D,T\}} \mathcal{S}_{i^{[k](r)},\alpha}\mathcal{T}_\alpha \mathcal{S}_{\alpha,i'^{[k'](r')}}=\mathcal{T}^{\dagger}_{i^{[k](r)}} \mathcal{S}_{i^{[k](r)},i'^{[k'](r')}} \mathcal{T}^{\dagger}_{i'^{[k'](r')}}
\end{equation}
where we only need to consider the sum $\alpha$ running over the diagonal and twisted primary labels, since $\Sc_{i^{[k](r)},[j_1\dots j_N]}=0$.
The sum over diagonal operator labels on the LHS is swiftly calculated to be:
\begin{equation}
      \sum_{j,s} \mathcal{S}_{i^{[k](r)},j^{(s)}}\mathcal{T}_{j^{(s)}} \mathcal{S}_{j^{(s)},i'^{[k'](r')}}=\frac{1}{N} \left(ST^NS\right)_{ii'}\delta_{k+k',0}
\end{equation}
and we use  (\ref{eq:P1}) to rewrite the above as:
\begin{equation}
       \sum_{j,s} \mathcal{S}_{i^{[k](r)},j^{(s)}}\mathcal{T}_{j^{(s)}} \mathcal{S}_{j^{(s)},i'^{[k'](r')}}=\frac{1}{N} \left(T^{-1/N}P_{-1}T^{-1/N}\right)_{ii'}\delta_{k+k',0}
\end{equation}
Next, we want to evaluate the sum over twisted operator labels:
\begin{equation}
  \sum_{j,s,t} \mathcal{S}_{i^{[k](r)},j^{[t](s)}}\mathcal{T}_{j^{[t](s)}} \mathcal{S}_{j^{[t](s)},i'^{[k'](r')}}=\frac{1}{N^2}\sum_{s,t} \omega^{-t(r+r')-s(k+k'-t)} (P_{tk^{-1}}T^{1/N}P_{k't^{-1}})_{ii'}
\end{equation}
We sum over $s$, to obtain a $\delta_{k+k',t}$ factor in the sum. If $k+k'=0 $, the expression vanishes since $t$ runs from 1 to $N-1$. So, we obtain:
\begin{equation}
     \sum_{j,s,t} \mathcal{S}_{i^{[k](r)},j^{[t](s)}}\mathcal{T}_{j^{[t](s)}} \mathcal{S}_{j^{[t](s)},i'^{[k'](r')}}=\frac{(1-\delta_{k+k',0})}{N} \omega^{-(k+k')(r+r')} (P_{(k+k')k^{-1}}T^{1/N}P_{1-k(k+k')^{-1}})_{ii'}
\end{equation}
Using the relation (\ref{eq:TPTPT}), the sum over twist fields becomes:
\begin{equation}
     \sum_{j,s,t} \mathcal{S}_{i^{[k](r)},j^{[t](s)}}\mathcal{T}_{j^{[t](s)}} \mathcal{S}_{j^{[t](s)},i'^{[k'](r')}}=\frac{(1-\delta_{k+k',0})}{N} \omega^{-(k+k')(r+r')} (T^{-1/N}P_{k'k^{-1}}T^{-1/N})_{ii'}
\end{equation}
so that  the LHS of (\ref{eq:TTbllockST3}) is just:
\begin{equation}
    \sum_{\alpha \in \{D,T\}} \mathcal{S}_{i^{[k](r)},\alpha}\mathcal{T}_\alpha \mathcal{S}_{\alpha,i'^{[k'](r')}}= \frac{1}{N} \omega^{-(k+k')(r+r')} (T^{-1/N}P_{k'k^{-1}}T^{-1/N})_{ii'}
\end{equation}
Substitution of our expressions for the modular data on the RHS of (\ref{eq:TTbllockST3}) completes the proof.

\bibliographystyle{unsrt}
\bibliography{references.bib}
\end{document}